\documentclass[a4paper,aps,prd,showpacs,preprintnumbers,twocolumn,nofootinbib,superscriptaddress]{revtex4-1}

\usepackage[english]{babel}

\usepackage[utf8x]{inputenc}
\usepackage[T1]{fontenc}
\usepackage{ucs}
\usepackage{microtype}
\usepackage{newtxtext,newtxmath}
\usepackage{color}

\usepackage{url}
\usepackage{hyperref}

\usepackage{xspace}
\usepackage{lastpage}
\usepackage{stmaryrd}

\usepackage{graphicx}
\usepackage{empheq}
\usepackage{ifthen}
\usepackage{tensor}
\usepackage{paralist}

\usepackage{amsmath}
\usepackage{amsfonts}
\usepackage{amssymb}
\usepackage{mathrsfs}
\usepackage{extarrows}
\usepackage{verbatim}
\usepackage{upgreek}
\usepackage{wasysym}

\usepackage{natbib}

\newcommand\ph{\ensuremath{\varphi}}
\newcommand\eps{\ensuremath{\varepsilon}}

\newcommand{\cc}{\text{c.c.}}
\newcommand{\cst}{\mathrm{cst}}

\newcommand\define{\equiv}

\newcommand\vect[1]{\boldsymbol{#1}}

\newcommand\ex[1]{\mathrm{e}^{#1}}
\renewcommand\i{\ensuremath{\mathrm{i}}}

\newcommand\e[1]{_{\text{#1}}}
\newcommand\h[1]{^{\text{#1}}}
\newcommand\U[1]{\:\mathrm{#1}}

\newcommand{\dd}{\mathrm{d}}

\newcommand{\ddf}[3][]{\frac{\dd^{#1} #2}{\dd {#3}^{#1}}}

\renewcommand\lim[2]{\underset{ #1 \rightarrow #2 }{ \mathrm{lim} } \,}

\newcommand{\delimiters}[4][]{
\ifthenelse{ \equal{#1}{1} }{  #2 #3 #4  }
					{ \ifthenelse{\equal{#1}{2}}{ \big#2 #3 \big#4 }
						{ \ifthenelse{\equal{#1}{3}}{ \Big#2 #3 \Big#4 }
							{ \ifthenelse{\equal{#1}{4}}{ \bigg#2 #3 \bigg#4 }
								{ \ifthenelse{\equal{#1}{5}}{ \Bigg#2 #3 \Bigg#4 }
									{ \left#2 #3 \right#4 }
								}
							}
						}
					}
													}

\newcommand{\pa}[2][]{\delimiters[#1]{(}{#2}{)}}
\newcommand{\pac}[2][]{\delimiters[#1]{[}{#2}{]}}

\newcommand{\ev}[2][]{\delimiters[#1]{\langle}{#2}{\rangle}}

\newcommand{\ket}[2][]{\delimiters[#1]{|}{#2}{\rangle}}

\newcommand{\Lagrangian}{\mathcal{L}}
\newcommand{\EOM}{\mathcal{E}}
\newcommand{\mass}{\mathcal{M}}
\newcommand{\amplitude}{\mathcal{A}}
\newcommand{\kinetic}{\mathcal{K}}
\newcommand{\region}{\mathcal{R}}
\newcommand{\propagator}{\mathcal{D}}

\def\l{\left}
\def\r{\right}

\begin{document}

\title{Horndeski and the Sirens}

\author{Charles Dalang}
\email{charles.dalang@unige.ch}
\affiliation{D\'{e}partement de Physique Th\'{e}orique, Universit\'{e} de Gen\`{e}ve,\\
24 quai Ernest-Ansermet, 1211 Gen\`{e}ve 4, Switzerland}

\author{Pierre Fleury}
\email{pierre.fleury@uam.es}
\affiliation{D\'{e}partement de Physique Th\'{e}orique, Universit\'{e} de Gen\`{e}ve,\\
24 quai Ernest-Ansermet, 1211 Gen\`{e}ve 4, Switzerland}
\affiliation{Instituto de F\'isica Te\'orica UAM-CSIC,
Universidad Auton\'oma de Madrid,\\
Cantoblanco, 28049 Madrid, Spain}

\author{Lucas Lombriser}
\email{lucas.lombriser@unige.ch}
\affiliation{D\'{e}partement de Physique Th\'{e}orique, Universit\'{e} de Gen\`{e}ve,\\
24 quai Ernest-Ansermet, 1211 Gen\`{e}ve 4, Switzerland}

\preprint{IFT-UAM/CSIC-168}

\begin{abstract}
Standard sirens have been proposed as probes of alternative theories of gravity, such as Horndeski models. Hitherto, all studies have been conducted on a homogeneous-isotropic cosmological background, which is unable to consistently account for realistic distributions of matter, and for inhomogeneities in the Horndeski scalar field. Yet, the latter are essential for screening mechanisms.
In this article, we comprehensively analyze the propagation of Horndeski gravitational waves in an arbitrary background spacetime and scalar field. We restrict our study to the class of theories in which gravitational waves propagate at light speed, we work in the geometric-optics regime, and we neglect scalar radiation.
We find that kinetic braiding produces a nonphysical longitudinal mode, whereas conformal coupling affects the amplitude of the standard transverse modes but not their polarization. We confirm that any observable deviation from general relativity depends on the local value of the effective Planck mass at emission and reception of the wave. This result is interpreted as the conservation of the number of gravitons.
\end{abstract}

\date{\today}
\maketitle

\section{Introduction: Horndeski's Odyssey}

\begin{flushright}
{\sl
\indent Tell me of a complicated theory.\\
\indent Urania, tell me how it wandered and was lost\\
\indent when it had attempted to replace the relativity of Einstein.\footnote{Adapted from the opening lines of Homer's Odyssey, in a recent translation by Emily Wilson~\cite{Odyssey}.}
}
\end{flushright}

\medskip

This theory, which more accurately is a crew of theories raised by Horndeski~\cite{Horndeski:1974wa}, supplements general relativity (hereafter GR) with a scalar degree of freedom. From a phenomenological point of view, Horndeski theories may be considered an attempt to model dark energy, i.e. the cause of the acceleration of cosmic expansion. From a more theoretical point of view, they could represent the effective behavior of more elaborate theories of gravity beyond GR~\cite{Dvali:2000hr, Nicolis:2008in, Chow:2009fm, Agarwal:2009gy, deRham:2011by}.

Because of their additional scalar, Horndeski theories predict a different gravitational strength compared to GR. The amplitude of such a difference, however, is highly constrained by the exquisite observation of our Solar System~\cite{Will:2014kxa}. In order to escape the Sun's scrutinizing eye, Horndeski nevertheless managed to disguise so as not to be noticed. Screening mechanisms can indeed cancel the effect of the scalar field when the local gravitational potential~\cite{Khoury:2003aq}, gravitational field~\cite{Babichev:2009ee}, or matter density~\cite{Vainshtein:1972sx} exceeds some threshold. In principle, this allows Horndeski theories to be indistinguishable from GR in the Solar System, while significantly differing from it on cosmological scales. See, however, Refs.~\cite{Babichev:2011iz, Wang:2012kj, Jimenez:2015bwa} for mitigation.

Other perilous challenges were nonetheless awaiting Horndeski, after he escaped the Solar System; he had to navigate through the vast universal ocean, avoiding rogue waves, traitorous cosmic flows, and the mighty creatures hidden in the cosmic web. It was indeed argued in Ref.~\cite{Creminelli_2020} that gravitational waves (hereafter GWs) could destabilize Horndeski's scalar field in the presence of kinetic braiding. Besides, many current and near-future tests of Horndeski theories rely on the observation of the local and large-scale structure of the Universe~\cite{Baker:2019gxo}, including galaxy clustering, weak gravitational lensing, and redshift-space distortions~\cite{Gleyzes:2015rua, Bellini:2015xja, Alonso:2016suf, Noller:2018wyv, Noller:2018eht}, but also relativistic effects~\cite{Bonvin:2018ckp} and more local observations~\cite{Desmond:2018euk}.

But of all the dangers that Horndeski had to face, the deadliest one was undoubtedly the spell cast by the \emph{circecular}\footnote{We ask the demanding reader to excuse us for this poetic license.} ballet of two neutron stars, which condemned the best members of his crew. By constraining the speed of GWs to be equal to the speed of light, the combined observation of GW170817~\cite{TheLIGOScientific:2017qsa} and GRB~170817~\cite{Goldstein:2017mmi, Savchenko:2017ffs} thereby ruled out a significant fraction of the parameter space of Horndeski's theories~\cite{Kimura:2011qn, McManus:2016kxu, Lombriser:2015sxa, Lombriser:2016yzn, Ezquiaga:2017ekz, Creminelli:2017sry, Sakstein:2017xjx, Langlois:2017, Baker:2017hug}. In particular, it has put a lot of pressure on the so-called self-accelerating models~\cite{Lombriser:2016yzn}, in which the cause of the acceleration of cosmic expansion may be attributed to a genuine gravitational effect, rather than some form of dark energy.

Only two members of Horndeski's crew survived the spell, namely conformal coupling and cubic Galileon; but the gods' wrath was still upon them, and that is how Horndeski met the sirens. Merging binary systems of compact objects, such as black holes or neutron stars, are efficient emitters of GWs. Since the measurement of their waveform directly gives access to their distance, such objects were cognominated \emph{standard sirens}~\cite{Holz:2005df}. Just like standard candles, standard sirens can be used to build a Hubble diagram, i.e. a measurement of the relation between distance and redshift across the Universe. As such, they also offer a way to measure today's rate of cosmic expansion
~\cite{1986Natur.323..310S,Chen:2017rfc,Soares-Santos:2019irc}. While candles and sirens produce the same Hubble diagram in GR, differences are expected in Horndeski theories, notably through conformal coupling. That is why standard sirens were recently argued to be a key probe of gravity model beyond GR~\cite{Saltas:2014dha, Lombriser:2015sxa, Amendola:2017ovw, Belgacem:2017ihm, Belgacem:2018lbp, Lagos:2019kds, DAgostino:2019hvh}, especially in the future era of the Laser Interferometer Space Antenna (LISA)~\cite{Lisa:2020, Belgacem:2019pkk}.

As noticed in Ref.~\cite{Amendola:2017ovw}, and further explored~\cite{Dalang:2019fma} by two of the authors of the present article, if the environments of both the siren and her observer are screened, which is more than likely, then the aforementioned difference between electromagnetic and GW Hubble diagrams may be suppressed, thereby questioning the relevance of that probe of alternative theories of gravity, such as Horndeski's. Specifically, it was argued that the distance measured with a standard siren, $D\e{G}$, differs from the standard electromagnetic luminosity distance~$D\e{L}$ as
\begin{equation}\label{eq:DG_DL_introduction}
    D\e{G} =  \frac{M\e{o}}{M\e{s}} \, D\e{L} \ ,
\end{equation}
where $M\e{s}$, $M\e{o}$ respectively denote the \emph{local} effective Planck mass\footnote{In this article, the effective Planck mass~$M$ will refer to the prefactor of Ricci curvature in the action of gravitation. It shall be distinguished from the notion of effective Newton's constant~$G\e{eff}$, which more commonly refers to the quantity involved in the Poisson equation. In general, $M^{-2}\not= 8\pi G\e{eff}$.} at the source and at the observer. Thus, if any screening mechanism enforces $M\e{s}=M\e{o}$, then $D\e{G}=D\e{L}$ and the GW and electromagnetic Hubble diagram coincide. If, on the contrary, $M\e{s}\not=M\e{o}$, then one must also account for that effect in the trigger of supernova explosions, which may also lead to a coincidence between the two Hubble diagrams. In other words, standard-siren tests of Horndeski theories seem to lie between Scylla and Charybdis. 

At this point, we stress that every rationale underlying standard-siren tests of gravity, including Eq.~\eqref{eq:DG_DL_introduction}, was hitherto made for GWs propagating on an ideal Friedmann-Lema\^{i}tre-Robertson-Walker (FLRW) background, which is to the real Universe what Homer's Odyssey is to Greek history; in particular, the FLRW background cannot consistently account for spatial variations of the Horndeski scalar field. Yet, allowing for such variations is crucial, e.g. to model screening mechanisms, but also to evaluate the impact of kinetic braiding, which couples GWs to derivatives of the scalar field.

The \emph{telos} of this article--besides making epic puns--is to fill this gap, by investigating the propagation of GWs in Horndeski theories, for arbitrary spacetime backgrounds, and arbitrary variations of the scalar field. After briefly presenting the Horndeski models in Sec.~\ref{sec:Horndeski}, we analyze in Sec.~\ref{sec:GWs} the propagation of GWs within the geometric optics regime; in particular, we demonstrate the general validity of Eq.~\eqref{eq:DG_DL_introduction} beyond the homogeneous and isotropic FLRW background. Finally, in Sec.~\ref{sec:graviton_number_conservation}, we show that this relation between GW and luminosity distances is related to the conservation of the number of gravitons. We summarize and conclude in Sec.~\ref{sec:conclusion}.

Throughout this article, unlike Pheidippides, Greek indices modestly run from $0$ to $3$; bold symbols indicate three-vectors; a comma denotes a partial derivative; and a semicolon denotes a covariant derivative associated with the Levi-Civita connection. Symmetrization and antisymmetrization of indices follow $X_{(\mu\nu)}\define \frac{1}{2}(X_{\mu\nu}+X_{\nu\mu})$ and $X_{[\mu\nu]}\define \frac{1}{2}(X_{\mu\nu}-X_{\nu\mu})$. We use the convention of Misner, Thorne and Wheeler~\cite{Misner:1974qy} for the metric signature and the Riemann tensor. Finally, we adopt units such that $c=1=\hbar$ except for Sec.~\ref{sec:graviton_number_conservation}.

\section{Horndeski's models of gravity}
\label{sec:Horndeski}

Horndeski theories~\cite{Horndeski:1974wa} form a class of extensions of GR involving an extra scalar degree of freedom. When it is nonminimally coupled to the spacetime geometry, this scalar can be seen either as a new physical interaction--a fifth force--or a modification of the laws of gravity.
Specifically, Horndeski's theories are the most general local theories involving the metric tensor~$g_{\mu\nu}$ of a four-dimensional spacetime Lorentzian manifold, coupled to a scalar field~$\ph$, and whose equations of motion deriving from an action principle are second order. This section briefly introduces the main equations of these theories, and reviews some of their properties.

\subsection{Horndeski's action}

In the so-called \emph{Jordan frame}, Horndeski's action is
\begin{equation}
\label{eq:HorndeskiAction}
S= S\e{g}[\ph, g_{\mu\nu}] + S\e{m}[\psi, g_{\mu\nu}] \ ,
\end{equation}
where $\psi$ refers to the matter fields of particle physics, assumed to be minimally coupled to the spacetime metric~$g_{\mu\nu}$ within their action~$S\e{m}$; it is not necessary to specify the expression of $S\e{m}$ for the purpose of this article. More importantly, $S\e{g}$ encodes the gravitational sector of the theory,
\begin{equation}
S\e{g}[\ph, g_{\mu\nu}]
= \sum_{i=2}^{5} S_i[\ph, g_{\mu\nu}]
= \frac{M\e{P}^2}{2} \int \dd^4 x \, \sqrt{-g} \;
    \sum_{i=2}^{5} \Lagrangian_i \ ,
\end{equation}
where $M\e{P}\define \sqrt{1/8\pi G}$ is the reduced Planck mass, $G$ being Newton's constant. The four Lagrangian densities~$\Lagrangian_i$ read
\begin{align}
\label{eq:Lagrangians}
\Lagrangian_2 &\define G_2(\ph, X) \ , \\ 
\Lagrangian_3 &\define G_3(\ph, X) \Box\ph \ ,\\
\Lagrangian_4 &\define G_4(\ph, X) R 
                + G_{4,X}(\ph, X) 
                    \pac{(\Box\ph)^2
                        -\ph_{;\mu\nu}\ph^{;\mu\nu}} \ ,\\
\Lagrangian_5 &\define G_5(\ph,X) E^{\mu\nu}\ph_{;\mu\nu}
                - \frac{1}{6} G_{5,X}(\ph,X)
                    \big[
                        (\Box\ph)^3 \nonumber\\
                        &\hspace{2cm}
                        - 3\Box\ph \, \ph^{;\mu\nu} \ph_{;\mu\nu}
                        + 2 \ph\indices{_;_\mu^\nu}
                            \ph\indices{_;_\nu^\rho}
                            \ph\indices{_;_\rho^\mu}
                    \big] \ .
\end{align}
In the above, we have introduced the following shorthand notation: $\Box\define g^{\mu\nu}\nabla_\mu\nabla_\nu$ is the D'Alembert operator, $X\define -\frac{1}{2}\ph^{,\mu}\ph_{,\mu}$, $R$ is the Ricci scalar, and $E_{\mu\nu}$ is the Einstein tensor.\footnote{We choose this notation instead of the more traditional~$G_{\mu\nu}$ because there are already many $G$s in these expressions.}

The Horndeski class encompasses all scalar-tensor theories, such as quintessence~\cite{Tsujikawa:2013fta}, Brans-Dicke models~\cite{Brans:1961sx}, $f(R)$ models~\cite{DeFelice:2010aj}, but also covariant Galileons~\cite{Deffayet:2009wt}. Any specific model is thereby determined by the four functions~$G_{2\ldots 5}(\ph, X)$, whose functional form is mostly free, apart from a few conditions ensuring stability and causality~\cite{Perenon:2015sla, Noller:2018wyv}. Note however that, for a given model, the functional form of the $G_{2\ldots 5}$ is not unique. One is, indeed, free to reparametrize the scalar field as $\ph\mapsto\Tilde{\ph}(\ph)$. One can also choose to reparametrize the metric through a so-called disformal transformation
\begin{equation}
g_{\mu\nu}
\mapsto
\tilde{g}_{\mu\nu}
= C(\ph) g_{\mu\nu} + D(\ph) \ph_{,\mu} \ph_{,\nu} \ ,
\end{equation}
where $C, D$ are two arbitrary functions of the scalar field. Such a transformation preserves the general form of Horndeski's action~\cite{Bettoni:2013diz}. It is then said that one works in a different \emph{frame}.\footnote{This concept of frame has nothing to do with its counterpart in mechanics or relativity; it is just a particular choice for the parametrization of the degrees of freedom $\ph, g_{\mu\nu}$ of the theory.}
Disformal transformations do not change the physical properties of the model--see e.g. Ref.~\cite{Francfort:2019ynz} for a detailed example with conformal transformations in cosmology--but they change the physical interpretation of the metric. Indeed, since $\tilde{g}_{\mu\nu}$ is, in general, nonminimally coupled to matter, it does not indicate the times and distances actually measured by an observer.

In this article, we choose to work in the Jordan frame, i.e. the parametrization such that the metric is minimally coupled to matter. This choice ensures that gravity acts on matter via~$g_{\mu\nu}$ only. In particular, interferometers designed to detect GWs, such as LIGO~\cite{Ligo:2020} and Virgo~\cite{Virgo:2020}, are then sensitive to waves of the metric field~$g_{\mu\nu}$ only. This does not mean that the scalar field has absolutely no effect on matter, but rather that it has to act on it via the geometry of spacetime.

The quasisimultaneous detection, in August 2017, of the GW-signal emitted by a neutron-star binary (GW170817~\cite{TheLIGOScientific:2017qsa}) and the associated $\gamma$-ray burst (GRB~170817~\cite{Goldstein:2017mmi, Savchenko:2017ffs}) proved that the propagation speed of GWs cannot differ from the speed of light by more than a part in $10^{15}$. Assuming that this difference is exactly zero, and excluding fine-tuned theories~\cite{Copeland:2018yuh}, viable Horndeski models must satisfy~\cite{Kimura:2011qn, McManus:2016kxu, Lombriser:2015sxa, Lombriser:2016yzn, Ezquiaga:2017ekz, Creminelli:2017sry, Sakstein:2017xjx, Langlois:2017, Baker:2017hug}
\begin{equation}
\label{eq:GWspeedconstraint}
G_{4,X} = G_5 = 0 \ .
\end{equation}

Two remarks may be formulated about Eq.~\eqref{eq:GWspeedconstraint}. First, the equality between GW speed and the speed of light has been experimentally tested for GWs of relatively small wavelengths (on the order of $10^6\U{m}$), i.e. relatively high energies, compared to, e.g., the relevant scales of cosmology. Thus, it may be cavalier to extrapolate this result and conclude that Eq.~\eqref{eq:GWspeedconstraint} holds at low energies~\cite{deRham:2018red}. Second, the neutron-star merger that produced the GW and GRB signal from which Eq.~\eqref{eq:GWspeedconstraint} was deduced, happened relatively close to us (about $40\U{Mpc}$ away). It is not excluded, in principle, that Eq.~\eqref{eq:GWspeedconstraint} only holds locally, but not at cosmological distances across our past lightcone \cite{Bonilla_2020}. Having pointed out these possible limitations, we shall nevertheless assume in this article that Eq.~\eqref{eq:GWspeedconstraint} does hold in the entire Universe, and is indeed a property of the theory.

\subsection{Equations of motion}

This section gives the equations of motion for $\ph, g_{\mu\nu}$ deriving from Horndeski's action with $G_{4,X}=G_5=0$. These will be the starting point of the present analysis.

\subsubsection{Metric tensor}

Let us start with what will eventually be our main focus, namely the equation of motion for the tensor field~$g_{\mu\nu}$. Varying Eq.~\eqref{eq:HorndeskiAction} with respect to $g^{\mu\nu}$, formally yields
\begin{equation}\label{eq:EoM_metric}
\sum_{i=2}^4 \EOM^i_{\mu\nu} = M\e{P}^{-2} \, T_{\mu\nu} \ ,
\qquad \text{with} \quad
\EOM^i_{\mu\nu}
\define \frac{2 M\e{P}^{-2}}{\sqrt{-g}} 
        \frac{\delta S_i}{\delta g^{\mu\nu}} \ ,
\end{equation}
and where $T_{\mu\nu}\define(-2/\sqrt{-g}) \delta S\e{m}/\delta g^{\mu\nu}$ is the matter energy-momentum tensor. The four pieces~$\EOM^{i=2\ldots 4}_{\mu\nu}$ of the equation of motion explicitly read
\begin{align}
\label{eq:E2}
\EOM^2_{\mu\nu} &= -\frac{1}{2}
                    \pa{
                        G_{2,X} \ph_{,\mu} \ph_{,\nu}
                        + G_2 g_{\mu\nu}
                        }\,, \\
\label{eq:E3}
\EOM^3_{\mu\nu} &= - G_{3,\ph} 
                    (\ph_{,\mu} \ph_{,\nu} + X g_{\mu\nu})
                    \nonumber\\
                    &\quad - G_{3,X}
                        \pac{
                            \ph_{,(\mu} X_{,\nu)}
                            + \frac{1}{2} \Box\ph \,
                                \ph_{,\mu} \ph_{,\nu}
                            - \frac{1}{2}\ph^{,\rho} X_{,\rho}     g_{\mu\nu}
                            }\,, \\
\label{eq:E4}
\EOM^4_{\mu\nu} &= G_4 E_{\mu\nu}
                    - G_{4;\mu\nu}
                    + \Box G_4 \, g_{\mu\nu} \ .
\end{align}
Recall that, in this article, $E_{\mu\nu}$ denotes the Einstein tensor.

Dividing the entire equation of motion~\eqref{eq:EoM_metric} by $G_4$, one sees that matter gravitates via the effective Planck mass
\begin{equation}
    M^2\define G_4 M\e{P}^2
        = \frac{G_4}{8\pi G}\ .
\end{equation}
Hence, the conformal factor~$G_4$ may be seen as a multiplicative correction to the Planck mass.

\subsubsection{Scalar field}

As mentioned earlier, we chose to work in the Jordan frame so as to maximally reduce interactions between matter and the scalar field. Thus, the dynamics of $\ph$ will be mostly irrelevant in this article. We nevertheless give its equation of motion below for completeness. Imposing that the variation of the action~\eqref{eq:HorndeskiAction} with respect to $\ph$ vanishes yields
\begin{multline}
\label{eq:EoM_scalar}
0 =
G_{2,\ph}
+G_{2,X} \Box\ph
- 2X G_{2,X\ph}
+ G_{2,XX} \ph^{,\mu} X_{,\mu} \\
+ \left(
        2 G_{3,\ph} - 2X G_{3,\ph X}
        + G_{3,XX} \ph^{,\mu} X_{,\mu}
        + G_{3,X} \Box\ph
    \right) \Box\ph \\
- 2X G_{3,\ph\ph}
+ 2G_{3,\ph X} \ph^{,\mu} X_{,\mu}
+ G_{3,XX} X_{,\mu} X^{,\mu}\\
- G_{3,X} (\ph^{;\mu\nu} \ph_{;\mu\nu}
           + R^{\mu\nu} \ph_{,\mu} \ph_{,\nu})
+ G_{4,\ph} R \ .
\end{multline}

Using Eq.~\eqref{eq:EoM_metric}, one can substitute the curvature terms $R_{\mu\nu}, R$ of Eq.~\eqref{eq:EoM_scalar} with their expression in terms of $\ph, T_{\mu\nu}$. The resulting equation being much longer than Eq.~\eqref{eq:EoM_scalar}, but not particularly illuminating, we shall not write it down explicitly. Nevertheless, it is interesting to point out that it contains terms proportional to $G_{3,X} T^{\mu\nu}\ph_{,\mu}\ph_{,\nu}$ and $G_{4,\ph} T$, where $T\define T^\mu_\mu$ is the trace of the matter energy-momentum tensor. Therefore, the scalar field is directly sourced by matter if either $G_{3,X}\not= 0$ (kinetic braiding~\cite{Deffayet:2010qz}) or $G_{4,\ph}\not=0$ (conformal coupling), even in the Jordan frame.

\subsection{Screening}

The presence of an extra scalar degree of freedom in Horndeski theories can lead to a rich gravitational phenomenology: variations of the effective Planck mass, violation of the equivalence principle~\cite{Hui:2009kc}, anomalous propagation of light, etc. However, such phenomena are tightly constrained in the Solar System and the Milky Way. Specifically, lunar laser ranging~\cite{Hofmann:2018myc} imposes that $|\dot{G}\e{eff}/G\e{eff}|< 10^{-14}\U{yr^{-1}}$, where $G\e{eff}$ denotes the gravitational coupling entering the effective Poisson equation. Besides, the 21-year monitoring~\cite{2015ApJ...809...41Z} of the timing of the pulsar binary PSR J1713+0747, about $1.2\U{kpc}$ away from us, imposes $|\dot{G}\e{eff}/G\e{eff}|< 10^{-12}\U{yr^{-1}}$ there. Regarding the effect of gravity on light propagation, and in particular the Shapiro time-delay effect, radio communication with the \textit{Cassini} spacecraft~\cite{Bertotti:2003rm} in solar conjunction, in 2002, constrained the $\gamma$ post-Newtonian parameter as $|\gamma-1|<2\times 10^{-5}$. Together with the constraint that GWs propagate at the speed of light, these should force gravitation to be extremely close to GR~\cite{Lombriser:2015sxa, Lombriser:2016yzn}.

Is there any way out? It is important to note that the conclusions of Refs.~\cite{Lombriser:2015sxa, Lombriser:2016yzn} rely on the assumption that Horndeski theories have the same behavior in the Solar System and on cosmic scales. However, it turns out that Horndeski theories generically possess screening mechanisms, which allow the scalar field to hide in situations relevant to, e.g., the Solar System. Conformally coupled scalar fields (which include $f(R)$ and symmetron models~\cite{PhysRevLett.104.231301}) can display the so-called chameleon mechanism~\cite{Khoury:2003aq, Hinterbichler:2011ca}, where $\ph$ is effectively suppressed when the gravitational potential is large enough. Another possible screening mechanism is the Vainstein effect, which was originally introduced in the context of massive gravity~\cite{Vainshtein:1972sx}, but is also relevant for Galileon-like models. In that case, self-interactions in the kinetic (rather than potential) part of the action of $\ph$ lead to a suppression of its effects when the matter density is large enough. Similarly, k-mouflage~\cite{Babichev:2009ee} operates when the gravitational acceleration is large enough.

Screening is not always a panacea. For instance, chameleon models cannot be screened in the Solar System and in the Milky Way while self-accelerating cosmic expansion (i.e. without the help of some form of dark energy)~\cite{Wang:2012kj}. Besides, the Vainshtein mechanism seems to be unable to screen cosmological time variations of the effective Planck mass, at least in Galileon models~\cite{Babichev:2011iz}; hence, such models cannot explain the cosmological dynamics without violating the stringent constraints on $|\dot{G}\e{eff}/G\e{eff}|$ or on $|\gamma-1|$.

\subsection{Absence of scalar waves?}
\label{subsec:scalar_waves}

In this article, \emph{we shall not consider waves in the scalar sector}. This may seem surprising at first sight. Indeed, as pointed out just after Eq.~\eqref{eq:EoM_scalar}, $\ph$ is effectively sourced by matter, so it must propagate waves just like the metric does. However GW sources, such as binary systems of black holes or neutron stars, are typically strong-field regions, which are very likely to be screened. Because screening suppresses the effects of the scalar field, it is natural to expect it to suppress scalar radiation as well. This intuition was confirmed both analytically~\cite{deRham:2012fw, Chu:2012kz} and numerically~\cite{Dar:2018dra} in the Vainshtein case.
From the observational point of view, the monitoring of the orbital period of the Hulse-Taylor binary pulsar has already set a strong upper bound on the energy loss that could be attributed to extra radiation modes~\cite{Will:2014kxa,Katsuragawa:2019uto}.

However, even in the absence of direct scalar radiation, some may be generated by "leakage" of tensor GWs in the scalar sector. Indeed, $\ph$ is sourced by Ricci curvature, and conversely curvature is sourced by $\Box\ph$; both may thus exchange energy as a GW propagates. This second-order derivative interaction may also affect the dispersion relation of both tensor and scalar waves. A comprehensive way of addressing this issue would consist in treating scalar and tensor perturbations simultaneously, and diagonalize the resulting system of propagation equations. Albeit relatively straightforward when the background spacetime and scalar field are trivial~\cite{Hou:2017bqj}, this analysis would be significantly more involved in general, and is left for future work. In what follows, the scalar field may thus be viewed as a kind of stiff medium, which will guide the propagation of GWs without actually being shaken by them.

%
%
%

Finally, one may object that the absence of scalar waves is a \emph{gauge-dependent} statement. Indeed, it can be expressed as follows: the typical scale over which $\ph=\bar{\ph}$ varies appreciably is much larger than the GW's wavelength. However, suppose that one performs a gauge transformation, i.e. an infinitesimal coordinate transformation~$x^\mu\mapsto x^\mu -\xi^\mu$, where $\xi^\mu$ varies as rapidly as the GW; then the scalar field becomes $\ph\mapsto\bar{\ph}+\delta\ph$, where $\delta\ph=\xi^\mu\bar{\ph}_{,\mu}$ behaves as a scalar wave. Therefore, to be specific, we will assume in this work that scalar waves are negligible \emph{in the harmonic gauge} (see Sec.~\ref{subsubsec:harmonic_gauge}). Although it follows a different rationale, the resulting setup is very similar to  the one considered in Ref.~\cite{Garoffolo:2019mna} (see Appendix~\ref{app:comparison_G19} for details).

\section{Gravitational waves in Horndeski models}
\label{sec:GWs}

The present section is the core of the article. We carefully analyze the propagation of GWs in the Horndeski theories described in Sec.~\ref{sec:Horndeski}, and the notion of distance measured with standard sirens. Contrary to previous studies, we do not restrict our discussion to a homogeneous-isotropic FLRW background spacetime, and work with an arbitrary background geometry and scalar field.

\subsection{Linearized equations of motion}

Let us assume the existence of a coordinate system such that the spacetime metric reads
\begin{equation}
\label{eq:metric_background_perturbation}
g_{\mu\nu} = \bar{g}_{\mu\nu} + h_{\mu\nu} \ ,
\qquad |h_{\mu\nu}| \ll 1 \ ,
\end{equation}
where $\bar{g}_{\mu\nu}$ represents the background spacetime, typically generated by astrophysical and cosmological sources, while the perturbation $h_{\mu\nu}$ stands for GWs. The equation of motion~\eqref{eq:EoM_metric} for $g_{\mu\nu}$ can then be formally linearized as
\begin{equation}
\sum_{i=2}^4
\EOM^i_{\mu\nu} =
\sum_{i=2}^4 \bar{\EOM}^i_{\mu\nu}
    + \delta\EOM^i_{\mu\nu}
= M\e{P}^{-2} \pa{\bar{T}_{\mu\nu} + \delta T_{\mu\nu}} \ ,
\end{equation}
where $\delta\EOM^{i}_{\mu\nu}, \delta T_{\mu\nu}$ are linear in $h_{\mu\nu}$ and its derivatives. By definition, the background metric satisfies $\sum_{i=2}^4 \bar{\EOM}^i_{\mu\nu}=M\e{P}^{-2}\bar{T}_{\mu\nu}$, and we are left with
\begin{equation}
\label{eq:linearized_EFE}
\sum_{i=2}^4
\delta\EOM^i_{\mu\nu}
- M\e{P}^{-2} \, \delta T_{\mu\nu}
= 0 \ .
\end{equation}
The term $\delta T_{\mu\nu}$ generally contains both traditional GW sources, such as coalescing binaries, and anisotropic stress, such as free-streaming neutrinos~\cite{Weinberg:2003ur, Scomparin:2019ziw}; only the latter influences the actual propagation of GWs. We shall neglect such sources throughout the article in order to focus on the effect of the scalar field on the GW propagation. Other contributions to $\delta T_{\mu\nu}$ may arise from perturbations of the metric in the background $T_{\mu\nu}$. Fortunately, for standard forms of matter such as perfect fluids, electromagnetic fields, and so on, $S\e{m}$ does not involve derivatives of the metric. Hence, $\delta T_{\mu\nu}$ is generally what we shall call a masslike term in the following. These have negligible impact on the propagation of GWs.

\subsubsection{Kinetic, damping, and mass terms}
\label{subsubsec:kinetic_damping_mass}

The linearized equation~\eqref{eq:linearized_EFE} is conveniently decomposed\footnote{The decomposition is not unique, since two covariant derivatives in $\kinetic_{\mu\nu}^i$ may be swapped at the price of introducing curvature terms, which add to the masslike terms $\mass_{\mu\nu}^i$.} into kinetic terms~$\kinetic^i_{\mu\nu}$, which contain second-order derivatives of $h_{\mu\nu}$; damping (or amplitude) terms~$\amplitude^i_{\mu\nu}$, with first-order derivatives; and masslike terms~$\mass^i_{\mu\nu}$ without any derivative: 
\begin{align}
\delta\EOM^2_{\mu\nu}
&= \mass^2_{\mu\nu} \ , \\
\delta\EOM^3_{\mu\nu}
&= \mass^3_{\mu\nu} + \amplitude^3_{\mu\nu} \ , \\
\delta\EOM^4_{\mu\nu}
&= \mass^4_{\mu\nu} 
    + \amplitude^4_{\mu\nu}
    + \kinetic^4_{\mu\nu} \ .
\end{align}

Let us briefly explain the origin of the relevant terms and give their expressions. First of all, the only kinetic term comes from the perturbation of the Einstein tensor; hence, it is proportional to the standard one obtained in GR,
\begin{equation}
\label{eq:kinetic_term_K4}
\kinetic^4_{\mu\nu}
= \frac{G_4}{2}
    \pac{
        2 \gamma\indices{_\rho_(_\mu _;^\rho _\nu_)}
        - \Box \gamma_{\mu\nu}
        - \gamma\indices{^\rho^\sigma_;_\rho_\sigma}    
            \bar{g}_{\mu\nu}
        } ,
\end{equation}
where $\gamma_{\mu\nu}$ denotes the usual trace-reversed metric perturbation
\begin{equation}
\gamma_{\mu\nu}
\define h_{\mu\nu} - \frac{1}{2} h \, \bar{g}_{\mu\nu} \ ,
\end{equation}
with $h\define h^\mu_\mu \define \bar{g}^{\mu\nu} h_{\mu\nu}$. Note that, in Eq.~\eqref{eq:kinetic_term_K4} and in all the remainder of this article, indices are raised and lowered by the \emph{background} metric~$\bar{g}_{\mu\nu}$; semicolons refer to background covariant derivatives~$\bar{\nabla}_\mu$, $\Box = \bar{g}^{\mu\nu}\bar{\nabla}_\mu\bar{\nabla}_\nu $.

The damping term associated with conformal coupling comes from the second covariant derivatives of $G_4$ in Eq.~\eqref{eq:E4}, because these feature products between $\partial G_4$ and Christoffel symbols. Its explicit expression is
\begin{equation}\label{eq:A_4_result}
\amplitude^4_{\mu\nu}
= \frac{1}{2} G_4^{,\rho}   
    (h_{\rho\mu;\nu}+h_{\rho\nu;\mu}-h_{\mu\nu;\rho})
    - G_{4,\rho} \gamma\indices{^\rho^\sigma_;_\sigma} \, \bar{g}_{\mu\nu} \,.
\end{equation}
Similarly, the cubic-Galileon piece leads to
\begin{multline}\label{eq:A_3_result}
\amplitude^3_{\mu\nu}
= \frac{1}{2} G_{3,X} \ph^{,\rho} \ph^{,\sigma}
    \bigg[
        \frac{1}{2} \ph^{,\lambda}
        h_{\rho\sigma;\lambda} \bar{g}_{\mu\nu}
        - h_{\rho\sigma;(\mu}\ph_{,\nu)}
    \bigg] \\
    + \frac{1}{2} G_{3,X} \ph_{,\sigma} \gamma\indices{^\rho^\sigma_;_\rho} \, \ph_{,\mu} \ph_{,\nu} \ .
\end{multline}
Note that Eqs.~\eqref{eq:A_4_result}, and \eqref{eq:A_3_result} feature both the metric perturbation~$h_{\mu\nu}$ and its trace-reversed counterpart~$\gamma_{\mu\nu}$; this choice was made to keep expressions as compact as possible.

The masslike terms~$\mass^i_{\mu\nu}$, which will be negligible in this work thanks to the eikonal approximation (see Sec.~\ref{subsec:eikonal}), are given in Appendix \ref{app:massterms} for completeness.

\subsubsection{Harmonic gauge}
\label{subsubsec:harmonic_gauge}

Just like in GR, the linearized equation of motion~\eqref{eq:linearized_EFE} enjoys a gauge invariance, because of the general covariance of Horndeski's action and of the resulting equations of motion. Recall that, under an infinitesimal coordinate transformation $x^\mu \mapsto x^\mu - \xi^\mu$, the metric perturbation changes as~\cite{Straumann:2013spu}
\begin{equation}
h_{\mu\nu} \mapsto h_{\mu\nu} + 2\xi_{(\mu;\nu)} \ ,
\end{equation}
which we shall refer to as a gauge transformation.

We can take advantage of the gauge invariance to simplify the equation of motion~\eqref{eq:linearized_EFE}. In particular, since under a gauge transformation
\begin{equation}
\label{eq:GaugeTransformation}
    \gamma\indices{_\mu_\nu^;^\nu} \mapsto \gamma\indices{_\mu_\nu^;^\nu} + \Box \xi_\mu + \bar{R}^\nu_\mu \xi_\nu \ ,
\end{equation}
it is always possible to impose
\begin{equation} \label{eq:HarmonicGauge}
\gamma\indices{_\mu_\nu^;^\nu} = 0 \ ,
\end{equation}
because if it were not the case initially, there would always exist a gauge field $\xi_\mu$ capable of ensuring it. Equation~\eqref{eq:HarmonicGauge} is known as the harmonic, or Hilbert, or De Donder gauge. It greatly simplifies the terms of interest as
\begin{align}
\label{eq:K4_harmonic}
\kinetic^4_{\mu\nu}
&= - \frac{1}{2} \, G_4 \, \Box\gamma_{\mu\nu}\,, \\
\label{eq:A4_harmonic}
\amplitude^4_{\mu\nu}
&= \frac{1}{2} G_4^{,\rho}   
    (h_{\rho\mu;\nu}+h_{\rho\nu;\mu}-h_{\mu\nu;\rho}) \,, \\
\label{eq:A3_harmonic}
\amplitude^3_{\mu\nu}
&= \frac{1}{2} G_{3,X} \ph^{,\rho} \ph^{,\sigma}
    \pac{
        \frac{1}{2} \ph^{,\lambda}
        h_{\rho\sigma;\lambda} \bar{g}_{\mu\nu}
        - h_{\rho\sigma;(\mu}\ph_{,\nu)}
        } \,.
\end{align}

\subsection{Eikonal approximation}
\label{subsec:eikonal}

The equation of motion for $h_{\mu\nu}$ may now be turned into a wave-propagation equation by working in the geometric-optics regime, i.e., using the eikonal approximation.

\subsubsection{Nature of the approximation}

In a nutshell, the eikonal approximation consists in assuming that the typical wavelength of the GW is much smaller than all the other characteristic length scales of the problem at hand. In that regime, the GW essentially behaves as a stream of particles, and all the phenomena related to its actual wave nature--interference, diffraction--can be neglected.

In more concrete terms, we shall introduce the ansatz
\begin{equation}\label{eq:ansatz}
h_{\mu\nu} = \frac{1}{2} \, H_{\mu\nu} \ex{\i w} + \cc
\end{equation}
where $H_{\mu\nu}$ represents the complex amplitude and polarization of the GW, $w$ its phase,\footnote{Although $\phi$ is a more common notation for a phase, $w$ has been chosen so as to avoid confusions with the scalar field.}${}^{,}$\footnote{Importantly, the ansatz~\eqref{eq:ansatz} assumes that the phase $w$ is common to all the components of $h_{\mu\nu}$. This does not mean that these components are all in phase, because $H_{\mu\nu}\in\mathbb{C}$, but it means that all the components of $h_{\mu\nu}$ must have the same wave vector. In particular, Eq.~\eqref{eq:ansatz} excludes gravitational birefringence. More generally, had the kinetic term (second-order derivatives) of $h_{\mu\nu}$ been effectively polarization dependent, the ansatz~\eqref{eq:ansatz} would not have sufficed. Such a polarization dependence generically appears in the presence of scalar waves in harmonic gauge.} and $\cc$ means complex conjugate. The eikonal approximation then implies that $w$ varies much faster than
\begin{enumerate}
    \item the GW amplitude,
    \item the background spacetime geometry, and
    \item the scalar field.
\end{enumerate}
If $\omega\sim\partial w$ denotes the typical cyclic frequency of the GW, the above translates into
\begin{equation}
\omega \gg H^{-1}\partial H \ , \quad
            \sqrt{|\bar{R}_{\mu\nu\rho\sigma}|} \ , \quad
            \ph^{-1}\partial\ph \ .
\end{equation}
In practice, $\omega^{-1}$ can be treated as a small parameter in terms of which one performs perturbative expansions; see e.g. Refs.~\cite{Misner:1974qy, Harte:2018wni} for further details.

For GWs falling in the LIGO maximum sensitivity, $\omega\sim 10^2\U{Hz}$, the typical wavelength is on the order of $10^6\U{m} \simeq 7\times 10^{-6}\U{AU}$, thereby making the eikonal approximation very accurate. For the expected LISA maximum sensitivity, $\omega\sim 10^{-3}\U{Hz}$, typical wavelengths approach the astronomical unit, making the eikonal approximation less applicable down to stellar scales. It remains, however, entirely valid for gravitational and scalar fields evolving on galactic scales, and a fortiori on cosmological scales. See Refs.~\cite{Harte:2018wni, Cusin:2019rmt} for recent attempts to model GWs beyond the geometric-optics regime.

\subsubsection{Propagation equations}

In the eikonal approximation, the various terms of the linearized equation of motion~\eqref{eq:linearized_EFE} for $h_{\mu\nu}$ obey a simple hierarchy
\begin{equation}
\underbrace{
            \kinetic^4_{\mu\nu}
            }_{\mathcal{O}(\omega^2)}
\gg
\underbrace{
            \amplitude^3_{\mu\nu} \,,
            \amplitude^4_{\mu\nu}
            }_{\mathcal{O}(\omega^1)}
\gg
\underbrace{
            \mass^i_{\mu\nu} \,,
            \delta T_{\mu\nu}
            }_{ \mathcal{O}(\omega^0)} \,,
\end{equation}
which leads us to directly neglect the masslike terms. Substituting Eqs.~\eqref{eq:K4_harmonic}-\eqref{eq:A3_harmonic} into the resulting $\kinetic^4_{\mu\nu}+\amplitude^4_{\mu\nu}+\amplitude^3_{\mu\nu}=0$, and subtracting its trace, we obtain the rather simple
\begin{equation}
\label{eq:propagation_equation_h_final}
G_4\Box h_{\mu\nu} = G_4^{,\rho} \pac{2 h_{\rho(\mu;\nu)} 
                                        - h_{\mu\nu;\rho}
                                    }
                    - G_{3,X} \ph^{,\rho} \ph^{,\sigma}
                        h_{\rho\sigma;(\mu} \ph_{,\nu)} \ .
\end{equation}

Using the wave-ansatz~\eqref{eq:ansatz} for $h_{\mu\nu}$, we have
\begin{align}
h_{\mu\nu;\rho}
&= \frac{1}{2}
    \pa{ H_{\mu\nu;\rho} 
        + \i k_\rho H_{\mu\nu} }
    \ex{\i w} + \cc \ ,\\
\Box h_{\mu\nu} &= \pa{ -k^\rho k_\rho H_{\mu\nu}
                        + \i\propagator H_{\mu\nu} 
                        + \Box H_{\mu\nu} }
                    \ex{\i w} + \cc \ ,
\end{align}
where $k_\mu\define w_{,\mu}$ denotes the wave four-vector associated with the eikonal~$w$, and $\propagator$ is a linear differential operator defined as
\begin{equation}
\label{eq:definition_propagator}
\propagator
\define 2 k^\mu\bar{\nabla}_\mu + k\indices{^\mu_;_\mu} \ .
\end{equation}
Equation~\eqref{eq:propagation_equation_h_final} may now be split into its real and imaginary parts.\footnote{More precisely, into cosine and sine components, due to the \cc\xspace terms.}
The real part is the \emph{dispersion relation},
\begin{equation}
\label{eq:dispersion_relation}
k^\mu k_\mu = 0 
\end{equation}
up to negligible masslike terms (including $\Box H_{\mu\nu}$), which indicates that GW wavefronts propagate at the speed of light. This is not surprising since we precisely restricted our study to this case via Eq.~\eqref{eq:GWspeedconstraint}. Note also that Eq.~\eqref{eq:dispersion_relation} implies the geodesic equation $k^\mu k_{\nu;\mu}=0$, because $k_\mu$ is a null gradient,\footnote{Taking the gradient of Eq.~\eqref{eq:dispersion_relation} leads to
\begin{equation*}
0
=(k^\mu k_\mu)_{;\nu}
= 2 k^\mu k_{\mu;\nu}
= 2 k^\mu w_{;\mu\nu}
= 2 k^\mu w_{;\nu\mu}
= 2 k^\mu k_{\nu;\mu} \ .
\end{equation*}
}
so that GWs propagate along null geodesics.

The imaginary part of Eq.~\eqref{eq:propagation_equation_h_final} governs the evolution of the GW amplitude, and reads
\begin{multline}
\label{eq:propagation_GW_amplitude}
G_4 \propagator H_{\mu\nu}
= 2 G_4^{,\rho} H_{\rho(\mu} k_{\nu)}
    - k^\rho G_{4,\rho} H_{\mu\nu} \\
    - G_{3,X} \ph^{,\rho}\ph^{,\sigma} H_{\rho\sigma} k_{(\mu} \ph_{,\nu)} \ .
\end{multline}
In GR, the above would reduce to $\propagator H_{\mu\nu}=0$, meaning that the GW energy would dilute as the area of the wavefront grows. Here, both conformal coupling ($G_{4,\ph}$) and kinetic braiding ($G_{3,X}$) seem to break this law. Note that the latter was absent from earlier studies, because in FLRW $\ph_{,\mu}=\ph_{,t}\delta^0_\mu$ and $H_{\mu\nu}=H_{ij}\delta^i_\mu \delta^j_\nu$, so that $\ph^{,\mu}H_{\mu\nu}=0$. We shall see, however, that this new term does not lead to observable corrections.

\subsection{Longitudinal and transverse modes}
\label{subsec:longitudinal_transverse_modes}

The GW amplitude $H_{\mu\nu}$ has six independent components ($10-4$ due to the harmonic-gauge condition). In GR, these can be reduced to two transverse modes using adequate gauge transformations. Things turn out to be more complicated in Horndeski theories. Namely, there will remain four longitudinal components in addition to the two traditional transverse ones. Although this longitudinal mode is nonphysical, and hence is not a proper degree of freedom, kinetic braiding seems to prevent us from gauging it away.

\subsubsection{Tetrad decomposition}
\label{subsubsec:tetrad_decomposition}

In order to decompose $H_{\mu\nu}$ into longitudinal and transverse components, it is useful to introduce a null tetrad\footnote{The choice of a null tetrad leads to compact expressions, but it may seem a bit obscure to readers unfamiliar with the associated formalism. See Appendix~\ref{app:null_tetrad} for further details, and relations with more traditional tetrads.} $(e^\alpha_\mu)\define (k_\mu, n_\mu, m_\mu, m_\mu^*)$, where $k_\mu$ is the wave four-vector, $m_\mu$ is complex, and a star denotes complex conjugation. This tetrad is defined with respect to the background metric; all its vectors are null, and the only nonvanishing scalar products are
\begin{equation} \label{eq:nonvanishingproducts}
\bar{g}^{\mu\nu} k_\mu n_\nu
=
\bar{g}^{\mu\nu} m_\mu m_\nu^*
= 1 \ .
\end{equation}
In terms of the null tetrad, the background metric reads
\begin{equation}
\bar{g}_{\mu\nu} = 2k_{(\mu} n_{\nu)} + 2 m_{(\mu} m^*_{\nu)} \ .
\end{equation}
While $k_\mu, n_\mu$ must be understood as being longitudinal with respect to the GW propagation, the complex vector $m_\mu$ and its conjugate are transverse to it. Finally, as $k_\mu$ is tangent to a null geodesic, it is parallel transported along itself. For convenience, we decide the other vectors~$n_\mu, m_\mu$ (and hence $m_\mu^*$) to be parallel transported as well,
\begin{equation}\label{eq:parallel-transported-tetrad}
k^\nu n_{\mu;\nu} = k^\nu m_{\mu;\nu} = 0 \ .
\end{equation}
This assumption does not restrict the generality of our analysis.

The harmonic gauge condition~\eqref{eq:HarmonicGauge}, in the eikonal approximation, reads
\begin{equation}\label{eq:harmonic_gauge_eikonal}
H_{\mu\nu} k^\nu = \frac{1}{2} H k_\mu \ .
\end{equation}
Inserting the tetrad decomposition of the GW amplitude as $H_{\mu\nu}=H_{\alpha\beta} e^\alpha_{(\mu} e^\beta_{\nu)}$, it is straightforward to check that Eq.~\eqref{eq:harmonic_gauge_eikonal} implies $H_{nn}=H_{nm}=H_{nm^*}=H_{mm^*}=0$. As a consequence, $H_{\mu\nu}$ can be generically written as the superposition of a longitudinal mode and a transverse-traceless mode:
\begin{equation}\label{eq:longitudinal_and_transverse}
H_{\mu\nu}
= H^{||}_{\mu\nu} + H^\perp_{\mu\nu} \ ,
\end{equation}
with
\begin{align}
\label{eq:longitudinal}
H^{||}_{\mu\nu}
&= 2 H_{(\mu} k_{\nu)} \ ,\\
\label{eq:transverse}
H^\perp_{\mu\nu}
&= H_\circlearrowleft m_{\mu} m_{\nu}
+ H_\circlearrowright m^*_\mu m^*_\nu \ .
\end{align}

In the transverse mode~$H_{\mu\nu}^\perp$, the two complex numbers $H_\circlearrowleft, H_\circlearrowright$ represent the amplitudes of the left and right circular polarizations of the GW; see Appendix~\ref{app:null_tetrad} for their relation with the plus and cross polarizations. If $H_\circlearrowleft=0$ or $H_\circlearrowright=0$, then the GW is circularly polarized; if $|H_\circlearrowleft|=|H_\circlearrowright|$, then the GW is linearly polarized; otherwise polarization is elliptical.

The longitudinal mode~$H_{\mu\nu}^{||}$ depends on a vector field~$H_{\mu}$, and hence carries four independent components. In GR, this mode can be entirely removed by a gauge transformation, provided that the energy-momentum tensor does not depend on derivatives of the metric, which is a reasonable restriction. In Horndeski theories, however, this operation can only be performed locally (see Appendix~\ref{app:gauge_removal_longitudinal}). Importantly, $H_{\mu\nu}^{||}$ should not be confused with the (physical) longitudinal mode that would be carried by scalar waves~\cite{Hou:2017bqj}, i.e. the mode that would produce tidal forces in the direction of the GW propagation.

\subsubsection{The longitudinal mode is nonphysical}
\label{subsubsec:longitudinal_nonphysical}

A good diagnostic of the physical content of a particular metric perturbation consists in computing the corresponding curvature perturbation. At linear order, the contribution of $h_{\mu\nu}$ to the Riemann tensor reads~\cite{Straumann:2013spu}
\begin{align}
\label{eq:perturbation_Riemann}
\delta R_{\mu\nu\rho\sigma}
&= \frac{1}{2} \pa{
                    h_{\mu\sigma;\nu\rho}
                    - h_{\mu\rho;\nu\sigma}
                    - h_{\nu\sigma;\mu\rho}
                    + h_{\nu\rho;\mu\sigma}
                } \\
&= k_{[\nu}H_{\mu][\rho} k_{\sigma]} \ex{\i w} + \cc \\
&= \frac{1}{2} \mathcal{R}_{\mu\nu\rho\sigma}\ex{\i w} + \cc
\ ,
\label{eq:curvature_amplitude}
\end{align}
where we only kept the leading $\mathcal{O}(\omega^2)$ terms, and we introduced the amplitude $\mathcal{R}_{\mu\nu\rho\sigma}\define 2k_{[\nu}H_{\mu][\rho} k_{\sigma]}$ of the curvature wave. This amplitude inherits the decomposition~\eqref{eq:longitudinal_and_transverse} of $H_{\mu\nu}$, that is, $\mathcal{R}_{\mu\nu\rho\sigma}=\mathcal{R}_{\mu\nu\rho\sigma}^{||}+\mathcal{R}_{\mu\nu\rho\sigma}^\perp$, with
\begin{align}
\mathcal{R}_{\mu\nu\rho\sigma}^{||}
&= 2k_{[\nu}H_{\mu]}k_{[\rho} k_{\sigma]}
    + 2k_{[\nu}H_{\rho]}k_{[\mu} k_{\sigma]} 
= 0 \ ,
\label{eq:no_curvature_longitudinal_mode}
\\
\mathcal{R}_{\mu\nu\rho\sigma}^\perp
&= 2 H_\circlearrowleft
    m_{[\mu} k_{\nu]} m_{[\rho} k_{\sigma]}
 + 2 H_\circlearrowright
    m^*_{[\mu} k_{\nu]} m^*_{[\rho} k_{\sigma]} \ .
\end{align}

In particular, Eq.~\eqref{eq:no_curvature_longitudinal_mode} shows that the longitudinal mode~$H_{\mu\nu}^{||}$ does not produce any curvature at leading order in the geometric optics regime, and hence it does not interact with matter. Furthermore, we shall see in Sec.~\ref{subsec:energy-momentum_GW} that this mode does not even carry energy-momentum in that regime. Thus, we shall consider it nonphysical just like in GR.

\subsubsection{Propagation of the transverse mode}

Projecting Eq.~\eqref{eq:propagation_GW_amplitude} successively on $m_{\mu} m_\nu$ and $m^*_\mu m^*_\nu$ yields the following propagation equation for the transverse amplitudes
\begin{equation}\label{eq:propagation_GW_transverse_amplitude}
\propagator \sqrt{G_4} H_\ocircle  = 0 \ ,
\end{equation}
where $H_\ocircle$ denotes either $H_\circlearrowleft$ or $H_\circlearrowright$ for short. Equation~\eqref{eq:propagation_GW_transverse_amplitude} holds regardless of the presence of a longitudinal mode~$H^{||}_{\mu\nu}$, which has no impact on the transverse modes~$H_\ocircle$. In order to interpret Eq.~\eqref{eq:propagation_GW_transverse_amplitude}, we shall first dedicate a few lines to the quantity~$k\indices{^\mu_;_\mu}$, which appears in the definition~\eqref{eq:definition_propagator} of~$\propagator$. For further details, we refer the curious reader to Sec.~2.3 of Ref.~\cite{Fleury:2015hgz}, about the Sachs optical scalars.

\begin{figure}[t]
    \centering
    \includegraphics[scale=1]{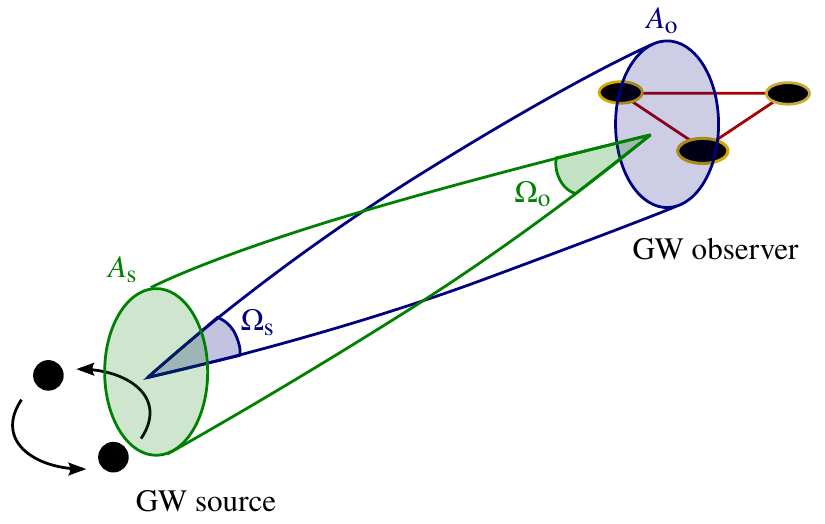}
    \caption{The natural setup involved in GW observations is such that wavefronts, and hence the associated geodesics, emerge from the source (blue). This is different from the setup involved in the definition of the angular-diameter distance (green).}
    \label{fig:Etherington}
\end{figure}

Consider a narrow bundle of null geodesics emerging from a GW source with a small solid angle $\Omega\e{s}$. This solid angle subtends a small patch of the wavefront, whose area $A$ grows as the wave propagates (see Fig.~\ref{fig:Etherington}). Let $\lambda$ be the affine parameter along the bundle such that $k^\mu=\dd x^\mu/\dd\lambda$. It can be shown that
\begin{equation}
    k\indices{^\mu_;_\mu} = \frac{1}{A} \ddf{A}{\lambda} \ ,
\end{equation}
which thus represents the local expansion rate of the wavefront. An observer who would measure the area~$A\e{o}$ of the bundle at their location could then define a notion of distance
\begin{equation}\label{eq:definition_D}
    D \define \sqrt{\frac{A\e{o}}{\Omega\e{s}}} \ ,
\end{equation}
so that $k\indices{^\mu_;_\mu} = 2\dd\ln D/\dd\lambda$. Equation~\eqref{eq:propagation_GW_transverse_amplitude} is then rewritten as
\begin{empheq}[box=\fbox]{equation}
\label{eq:propagation_GW_transverse_amplitude_D}
\ddf{}{\lambda}
\pac{ \sqrt{G_4(\ph)} \, D\,H_\ocircle }
= 0 \ ,
\end{empheq}
which shows that the GW amplitude evolves like $1/\sqrt{G_4}D$ as the wave propagates.

Quite importantly, $D$ as defined in Eq.~\eqref{eq:definition_D} is \emph{not} the usual angular-diameter distance~$D\e{A}\define\sqrt{A\e{s}/\Omega\e{o}}$, because the roles of the source and the observer have been swapped. These are nevertheless related by Etherington's reciprocity relation~\cite{1933PMag...15..761E}
\begin{equation}\label{eq:Etherington}
D = (1+z)\,D\e{A} \ ,
\end{equation}
where the $(1+z)$ factor comes from the fact that the solid angles $\Omega\e{o}, \Omega\e{s}$ are subject to different relativistic aberration effects.

Equation~\eqref{eq:propagation_GW_transverse_amplitude_D} is the main result of this article. It shows that in Horndeski theories, the physical amplitude of a GW decays as $1/(\sqrt{G_4}D)$ as the wave propagates. This result holds in \emph{any background spacetime} as long as the geometric-optics limit can be applied. It encapsulates all the relativistic effects that may affect $D$, such as Doppler beaming, gravitational redshift, lensing, etc.

\subsubsection{Polarization is parallel transported}
\label{subsubsec:parallel_transport}

Equation~\eqref{eq:propagation_GW_transverse_amplitude_D} also shows that the GW polarization is parallel transported along its worldline, just like in GR. This happens because the two circular modes~$H_\ocircle$ evolve identically and independently. Let us be more specific; the polarization tensor may be defined from the physical, transverse modes as
\begin{equation}
e_{\mu\nu}
\define
\frac{H_{\mu\nu}^\perp}{||H_\perp||}
=
\frac{H_\circlearrowleft e_{\mu\nu}^\circlearrowleft 
    + H_\circlearrowright e_{\mu\nu}^\circlearrowright}
{\sqrt{|H_\circlearrowleft|^2 + |H_\circlearrowright|^2}} \ ,
\end{equation}
with the left and right polarization basis tensors
\begin{equation}
e_{\mu\nu}^\circlearrowleft
\define m_\mu m_\nu \ ,
\qquad
e_{\mu\nu}^\circlearrowright
\define m^*_\mu m^*_\nu \ .
\end{equation}
Since our null tetrad is parallel transported along the GW geodesic, $e_{\mu\nu}^\circlearrowleft$ and $e_{\mu\nu}^\circlearrowright$ are parallel transported as well. They form an orthonormal set in the sense of the inner product~$\star$ defined in Appendix~\ref{app:null_product}. The norm $||H_\perp||^2 \define H^{\mu\nu}_\perp \star H^\perp_{\mu\nu}=|H_\circlearrowleft|^2 + |H_\circlearrowright|^2$ is also associated with that inner product. The polarization tensor $e_{\mu\nu}$ is normalized by definition. One may alternatively write $e_{\mu\nu}$ in terms of the plus and cross polarizations using Eqs.~\eqref{eq:left_plus_cross} and \eqref{eq:right_plus_cross}.

It is straightforward to check that Eq.~\eqref{eq:propagation_GW_transverse_amplitude_D} implies that $e_{\mu\nu}$ is parallel transported along the GW's geodesic, i.e.
\begin{equation}
k^\rho e_{\mu\nu;\rho} = 0 \ .
\end{equation}
Note that it is essential to define the GW polarization from its transverse modes~$H^\perp_{\mu\nu}$ only, otherwise one may find spurious departures from parallel transport~\cite{Garoffolo:2019mna}. This is because the nonphysical longitudinal mode~$H_{\mu\nu}^{||}$ is not parallel transported. We refer the interested reader to Appendix~\ref{app:comparison_G19} for more details.

For the sake of completeness, and to further justify the above, one may alternatively define the GW polarization directly from the perturbation of spacetime curvature. Albeit less standard, this approach has the advantage of involving only gauge-invariant and observable quantities. It is analogous to defining the polarization of an electromagnetic wave from its Faraday tensor~$F_{\mu\nu}$ rather than from its four-potential~$A_\mu$~\cite{1992grle.book.....S}.

We define the curvature polarization tensor as
\begin{equation}
e_{\mu\nu\rho\sigma}
\define \frac{\mathcal{R}_{\mu\nu\rho\sigma}}{||\mathcal{R}||}
= \frac{
    H_\circlearrowleft e^\circlearrowleft_{\mu\nu\rho\sigma}
+ H_\circlearrowright e^\circlearrowright_{\mu\nu\rho\sigma}
    }
    {\sqrt{|H_\circlearrowleft|^2 + |H_\circlearrowright|^2}}
    \ ,
\end{equation}
where $\mathcal{R}_{\mu\nu\rho\sigma}=\mathcal{R}^\perp_{\mu\nu\rho\sigma}$ is the amplitude of the curvature perturbation introduced in Eq.~\eqref{eq:curvature_amplitude}. The basis over which polarization is decomposed reads
\begin{align}
e^\circlearrowleft_{\mu\nu\rho\sigma}
&= 2 m_{[\mu} k_{\nu]} m_{[\rho} k_{\sigma]} \ ,
\\
e^\circlearrowright_{\mu\nu\rho\sigma}
&= 2 m^*_{[\mu} k_{\nu]} m^*_{[\rho} k_{\sigma]} \ .
\end{align}
Just like their rank-2 analogs, these two tensors are parallel transported; they form an orthonormal set in the sense of the inner product of Appendix~\ref{app:null_product}; finally, the normalization term reads $||\mathcal{R}||^2\define \mathcal{R}^{\mu\nu\rho\sigma} \star \mathcal{R}_{\mu\nu\rho\sigma}=|H_\circlearrowleft|^2+|H_\circlearrowright|^2$.

Again, it is straightforward to show that
\begin{equation}
k^\tau e_{\mu\nu\rho\sigma;\tau} = 0
\end{equation}
from Eq.~\eqref{eq:propagation_GW_transverse_amplitude_D},
which concludes our derivation of the parallel transport of the GW polarization.

\subsection{Consequences for standard sirens}

Merging binary systems, which are key sources of GWs, were nicknamed \emph{standard sirens} because the precise measurement of their waveform gives direct access to their distance~\cite{Holz:2005df}. However, since the very notion of distance is quite ambiguous in four dimensions~\cite{Fleury:2015hgz}, one may wonder whether, and how, this \emph{gravitational distance}~$D\e{G}$ is related to other known definitions of distance. For that purpose, one may consider how $D\e{G}$ is extracted from observations. As such, it is defined as the quantity that normalizes the GW amplitude. In terms of the plus and cross polarizations, one has indeed~\cite{Holz:2005df}
\begin{align}
h_+
&= \frac{[2\mathcal{M}_z^5\omega\e{o}^2(t)]^{1/3}}
        {D\e{G}}
    \, \pac{ 1+\cos^2\iota }
    \cos w(t) \ , \\
h_\times
&= \frac{[2\mathcal{M}_z^5\omega\e{o}^2(t)]^{1/3}}
        {D\e{G}}
    \, 2\cos\iota \,
    \sin w(t) \ ,
\end{align}
or, in terms of the amplitude of circular polarizations,
\begin{equation}
H_\ocircle
= \frac{[2\mathcal{M}_z^5\omega\e{o}^2(t)]^{1/3}}{D\e{G}} \,
     (1\pm\cos\iota)^2 ,
\end{equation}
where $\mathcal{M}_z=(1+z)\mathcal{M}$, is the redshifted chirp mass\footnote{This quantity can be accessed to via the accurate measurement of the time evolution of the GW frequency~\cite{Maggiore:1900zz}.} of the binary system producing the GW, $\omega\e{o}(t)$ is the observed GW cyclic frequency, and $\iota$ is the inclination angle formed between the line of sight and the orbital angular momentum of the binary. Since the observed frequency is related to the emitted one via $\omega\e{o}=\omega\e{s}/(1+z)$, we conclude that
\begin{equation}\label{eq:H_and_DG}
H_\ocircle \propto \frac{1+z}{D\e{G}} \ ,
\end{equation}
where the proportionality factor only depends on the properties of the source.
Therefore, combining Eq.~\eqref{eq:H_and_DG} with Eq.~\eqref{eq:propagation_GW_transverse_amplitude_D}, it appears that the gravitational distance reads
\begin{equation}
\label{eq:duality_distance_GW}
D\e{G}
= \sqrt{\frac{G_4(\ph\e{o})}{G_4(\ph\e{s})}} \, (1+z)^2 D\e{A} 
= \sqrt{\frac{G_4(\ph\e{o})}{G_4(\ph\e{s})}} \, D\e{L} \ ,
\end{equation}
where $\ph\e{s}$ and $\ph\e{o}$ are the values of the scalar field at emission and reception, respectively. In the last equality, we have introduced the standard electromagnetic luminosity distance, which is related to the angular distance by the distance-duality law $D\e{L}=(1+z)^2D\e{A}$ if the number of photons is conserved between the source and the observer.

We stress that although Eq.~\eqref{eq:duality_distance_GW} has been already widely used in the literature~\cite{Lombriser:2015sxa, Amendola:2017ovw, Belgacem:2017ihm, Belgacem:2018lbp, Lagos:2019kds, Belgacem:2019pkk, Dalang:2019fma, Wolf:2019hun}, it had actually never been rigorously derived in the general context of a nonhomogeneous Universe and an arbitrary distribution for the scalar field. Our result thus justifies the use of Eq.~\eqref{eq:duality_distance_GW} even in nonlinear setups, such as when the GW propagates through screened regions, as long as the eikonal approximation holds.

The importance of the fact that the gravitational distance~$D\e{G}$ only differs from the electromagnetic luminosity distance~$D\e{L}$ by the ratio of the \textit{local} values of $G_4(\ph)$ at emission and reception, was recently emphasized by Ref.~\cite{Dalang:2019fma}. This was shown to jeopardize standard-siren tests of modified gravity. On the one hand, if screening completely washes out the effect of $G_4$ in high-density regions, thereby forcing $G_4(\ph\e{o})=G_4(\ph\e{s})$, then $D\e{G}= D\e{L}$ and hence the modification of gravity remains as hidden as Odysseus' men inside the Trojan horse. On the other hand, if $G_4(\ph\e{o})\not=G_4(\ph\e{s})$, then the resulting modification of the effective Planck mass also affects the explosion mechanism of type Ia supernovae. This effectively changes their observed luminosity distance in a way that potentially cancels\footnote{The effectiveness of this cancellation depends on the exact relation between the supernova peak luminosity and the Chandrasekhar mass.} the difference between $D\e{L}$ and $D\e{G}$.

More generally, we stress that in the case of theories where variations of the effective Planck mass are a universal time evolution, standard-siren tests cannot compete with the combination of lunar laser ranging, Shapiro time delay, and GW propagation speed. These lead to constraints whose precision beats LISA forecasts by four orders of magnitude~\cite{Tsujikawa:2019pih}. Having said that, because of Eq.~\eqref{eq:duality_distance_GW}, standard sirens are still relevant to constrain scenarios in which the Planck mass would be constant in time, but spatially inhomogeneous.

\section{Physical interpretation: graviton-number conservation}
\label{sec:graviton_number_conservation}

In GR, the relation between luminosity distance and angular-diameter distance can be understood as a consequence of the conservation of photon number. In this section, we demonstrate that the relation~\eqref{eq:duality_distance_GW} between $D\e{G}$ and $D\e{A}$ can be interpreted as graviton-number conservation. This is true in GR but also in Horndeski models, where the concept of energy of a GW, and of a graviton, must be examined with care.

\subsection{Energy-momentum of a gravitational wave}
\label{subsec:energy-momentum_GW}

As a first step, it is instructive to determine the energy-momentum carried by a GW in Horndeski theories. In particular, we aim to determine the impact of the couplings~$G_3, G_4$ to the scalar field. This paragraph essentially follows the logic of Ref.~\cite{Straumann:2013spu}, adapting it from GR to Horndeski models.

By analogy with any form of matter, one defines the energy-momentum of a GW through its ability to \emph{gravitate}, i.e. to generate spacetime curvature. More precisely, the energy-momentum tensor~$T_{\mu\nu}\h{GW}$ of a GW quantifies the backreaction of the metric perturbation~$h_{\mu\nu}$ on the background metric~$\bar{g}_{\mu\nu}$. In order to evaluate it, one must expand the tensor equation of motion~\eqref{eq:EoM_metric} at second order in $h_{\mu\nu}$. This formally reads
\begin{equation}
\sum_{i=2}^4
\bar{\EOM}^i_{\mu\nu}
+ \delta^{1}\EOM^i_{\mu\nu}
+ \delta^{2}\EOM^i_{\mu\nu}
= M\e{P}^{-2}
    \pa[2]{ \bar{T}_{\mu\nu}
        + \delta^{1}T_{\mu\nu}
        + \delta^{2}T_{\mu\nu} } ,
\end{equation}
where a superscript on $\delta$ indicates the expansion order in $h_{\mu\nu}$. By construction, the first-order terms cancel. The second-order terms being quadratic in $h_{\mu\nu}$ and its derivatives are highly oscillating about a generically nonzero mean. Hence, one can split any such term as
\begin{equation}
\label{eq:split_average_oscillatory}
\delta^2\EOM_{\mu\nu} =
\ev{ \delta^2\EOM_{\mu\nu} }
+ \delta^2\EOM_{\mu\nu}\h{osc} ,
\end{equation}
where the brackets must be understood as a spacetime average over a region that is typically much larger than the GW wavelength, but much smaller than the other relevant length scales of the problem. On the right-hand side of Eq.~\eqref{eq:split_average_oscillatory}, the second term has zero average and is rapidly varying. On the contrary, the first term evolves on much larger scales, and hence is naturally thought of as a small correction to the background equation of motion:
\begin{equation}
\sum_{i=2}^4
\bar{\EOM}^i_{\mu\nu}
= M\e{P}^{-2}
    \pa[2]{ \bar{T}_{\mu\nu}
            + T_{\mu\nu}\h{GW} } ,
\end{equation}
where we defined
\begin{equation}
\label{eq:energy-momentum_GW_def}
T_{\mu\nu}\h{GW}
\define \ev{\delta^2 T_{\mu\nu}} 
        - M\e{P}^2 
        \sum_{i=2}^4\ev{\delta^2\EOM^i_{\mu\nu}} \ .
\end{equation}
This is how a GW gravitates.

As such, $T_{\mu\nu}\h{GW}$ is quadratic in $h_{\mu\nu}$ and its derivatives. In fact, examining Eqs~\eqref{eq:E2}, \eqref{eq:E3}, \eqref{eq:E4}, it is easy to see that any quadratic term must take either of the following four forms: $h^2, h\partial h, (\partial h)^2, h\partial^2 h$. The terms with only one derivative of $h$ vanish on average, because they are proportional to $\sin w \cos w$. The terms with two derivatives then completely overcome the terms with no derivatives, since
\begin{equation}
    (\partial h)^2, h\partial^2 h \sim \omega^2 h^2 \ ,
\end{equation}
and $\omega$ is very large. Hence, we shall only keep those in the expression of $T\h{GW}_{\mu\nu}$. It is straightforward to check that such terms can only come from $\EOM^4_{\mu\nu}$ and more precisely from the Einstein tensor. As a result,
\begin{equation}
T_{\mu\nu}\h{GW}
= - G_4 M\e{P}^2 \ev{\delta^2 E_{\mu\nu}}
= - M^2 \ev{\delta^2 E_{\mu\nu}} \ .
\end{equation}
In other words, the only difference with GR seems to be the conformal factor~$G_4$. However, before applying standard results too quickly, we must remember that the cubic coupling~$G_{3}$ generates a longitudinal mode~$h^{||}_{\mu\nu}$. This prevented us, in particular, to impose the usual transverse-traceless gauge, in which the rest of the calculation is usually performed.

Fortunately, it turns out that the longitudinal mode does not change the final result for $T_{\mu\nu}\h{GW}$. Let us prove this point. In harmonic gauge, after several integrations by parts, we find
\begin{equation}
\ev{\delta^2 E_{\mu\nu}}
= -\frac{1}{4}
    \ev{
        h_{\rho\sigma;\mu} h\indices{^\rho ^\sigma _{;\nu}}
        - \frac{1}{2} h_{,\mu} h_{,\nu}
        } ,
\end{equation}
in which it can be explicitly checked that the longitudinal mode~$h_{\mu\nu}^{||}$ does not contribute. This mode carrying neither curvature (see Sec.~\ref{subsubsec:longitudinal_nonphysical}) nor energy-momentum, we conclude that it is entirely nonphysical in the eikonal regime. Nevertheless, the apparent impossibility to eliminate that mode by gauge transformations indicates that it may play a role beyond geometric optics~\cite{Harte:2018wni, Cusin:2019rmt}.

The final result, in terms of the transverse amplitude, reads
\begin{equation}
\label{eq:energy_momentum_tensor_GW_final}
T_{\mu\nu}\h{GW}
= \frac{1}{8}\,M^2 ||H_\perp||^2 k_\mu k_\nu\,.
\end{equation}
with $||H_\perp||^2=|H_\circlearrowleft|^2+|H_\circlearrowright|^2$.
This energy-momentum tensor differs from GR only through the presence of the effective Planck mass~$M$, and hence it shares most of its properties.

\subsection{Energy-momentum of a graviton}

In order to proceed towards the definition of the number of gravitons in a GW, we must clarify what we may call a graviton. The definition that we adopt here relies on energetic properties: a graviton will be a quantum of gravitational energy, just like a photon is a quantum of electromagnetic energy. In technical terms, it will thus be an eigenstate of the Hamiltonian operator. As such, it is natural to expect--at least in GR--that the four-momentum of a single graviton is related to its wave vector as $p^\mu=\hbar k^\mu$, where $\hbar$ is the reduced Planck constant.\footnote{From now on, we restore $\hbar$ in order to keep track of quantum properties.} Does that picture change in Horndeski theories? In particular, does the conformal factor~$G_4(\ph)$ change the energy of a graviton?

A heuristic argument suggests a negative answer to the above question, i.e. $p^\mu=\hbar k^\mu$ even in the presence of $G_4(\ph)$. The conformal factor may be understood as a variation of the effective Planck mass in space and time. But since $M\e{P}$ is not involved in $p^\mu=\hbar k^\mu$, why should its variation affect that relation at all? In electromagnetism, an inhomogeneous dielectric medium is essentially equivalent to vacuum with an inhomogeneous permittivity~$\eps=n^2\eps_0$, where $n$ is the optical index of the medium~\cite{1998clel.book.....J}. Yet, such a modification does not change that $p^\mu=\hbar k^\mu$ for photons. We do not expect things to be different in gravitation. The remainder of this subsection is an attempt to further justify the above.

\subsubsection{Effective action}

Consider the classical behavior of the transverse modes in the geometric optics regime. The left-handed and right-handed modes~$h_\circlearrowleft, h_\circlearrowright$, defined as the projection over $m_\mu m_\nu$ and $m_\mu^* m_\nu^*$, respectively, evolve independently, just like two scalar fields. We also know that their typical wavelength is much shorter than all the other distance scales of the problem.

Let $\region$ be a region of spacetime that is much larger than the typical wavelength of the GWs under study, but smaller than the typical scale over which $\ph$ varies appreciably, and much smaller than the curvature radius of the background spacetime geometry. Thanks to the second condition, we can work in a normal coordinate system $x^\alpha$ in which the background metric is almost Minkowskian, $\bar{g}_{\alpha\beta}\approx \eta_{\alpha\beta}$ across $\mathcal{R}$.

An effective action for this system, leading to the correct equation of motion for the transverse mode $h_{\mu\nu}^\perp$, has the form $S\e{eff}[h_\circlearrowleft]+S\e{eff}[h_\circlearrowright]$, with
\begin{equation}
\label{eq:effective_action}
S\e{eff}[h_\ocircle] = -\frac{1}{32 \pi G} \int_{\region}
\dd^4 x \; G_4(\ph) \, h^\ocircle_{,\alpha} 
                        h_\ocircle^{,\alpha} \ .
\end{equation}
In Eq.~\eqref{eq:effective_action},  $h_\ocircle\define(1/2)H_\ocircle \ex{\i w}+\cc$, and, as before, $H_\ocircle$ refers to either of the independent polarizations $H_\circlearrowleft, H_\circlearrowright$. It is understood that, when varying the action, one has to impose $\delta h_\ocircle=0$ on $\partial\region$. It is then straightforward to check that $\delta S\e{eff}/\delta h_\ocircle=0$ implies $G_4 \Box h_\ocircle + G_{4,\alpha} h_\ocircle^{,\alpha}=0$, which is equivalent to the transverse part of Eq.~\eqref{eq:propagation_equation_h_final}.

At this point it is useful to define the canonical field
\begin{equation}
\label{eq:definition_chi}
\chi \define \sqrt{\frac{G_4(\ph)}{16 \pi G}}\, h_\ocircle \ ,
\end{equation}
which may be interpreted as a normalized version of the metric perturbation in the Einstein frame. In terms of this variable, the effective action is found to read
\begin{equation}
S\e{eff}[\chi] = -\frac{1}{2} \int_{\region} \dd^4 x
            \pac{ \chi_{,\alpha} \chi^{,\alpha}
                    + m^2 \chi^2} \ ,
\end{equation}
with the effective mass~$m^2=\Box \sqrt{G_4}/\sqrt{G_4}$. By definition, this mass is almost constant across $\region$; furthermore, it is of the same order of magnitude as the masslike terms that we have neglected so far thanks to the eikonal approximation. We can thus safely neglect it as well in what follows.

\subsubsection{Quantization}

Since $m^2$ can be considered constant across $\region$, or even neglected, the field $\chi$ is immediately quantized as:\footnote{For consistency, the range of $k$ over which we integrate must be such that $1/k$ remains much smaller than the size of $\region$.}
\begin{equation}
\label{eq:quantization_chi}
\hat{\chi}(x^\alpha)
= \int \frac{\dd^3 \vect{k}}{(2\pi)^3} \; \sqrt{\frac{\hbar}{2\omega(\vect{k})}}
    \pa{
        \hat{a}_{\vect{k}} \ex{\i k_\alpha x^\alpha}
        + \hat{a}_{\vect{k}}^\dagger 
                            \ex{-\i k_\alpha x^\alpha}
        } ,
\end{equation}
with $\omega^2(\vect{k})=\vect{k}^2+m^2\approx \vect{k}^2$, and $\hat{a}_{\vect{k}}, \hat{a}^\dagger_{\vect{k}}$ are the usual annihilation and creation operators for $\chi$, with commutators
\begin{equation}
[\hat{a}_{\vect{k}}, \hat{a}_{\vect{k}'}^\dagger]
= (2\pi)^3\delta(\vect{k}-\vect{k}') \ ,
\qquad
[\hat{a}_{\vect{k}}, \hat{a}_{\vect{k}'}]
= [\hat{a}_{\vect{k}}^\dagger, \hat{a}_{\vect{k}'}^\dagger]
= 0 \ ,
\end{equation}
so as to ensure canonical quantization.\footnote{Note in particular that the $\sqrt{\hbar}$ in Eq.~\eqref{eq:quantization_chi} ensures that the commutation relation between $\hat{\chi}$ and its conjugate momentum $\hat{\pi}=\partial_t\hat{\chi}$ reads $[\hat{\chi}(t,\vect{x}),\hat{\pi}(t,\vect{y})]=\i\hbar\delta(\vect{x}-\vect{y})$.}

The vacuum quantum state $\ket{0}$ is such that $\hat{a}_{\vect{k}}\ket{0}=0$, while the application of $\hat{a}^\dagger_{\vect{k}}$ populates any state with a quantum of $\chi$ with wave vector $\vect{k}$. Because of the proportionality~\eqref{eq:definition_chi} between $\chi$ and $h$, these states also describe quantum states of the original metric perturbation~$h$. It turns out that they are also quanta of energy, as shown in the next paragraph.

\subsubsection{Quanta of energy-momentum}

The energy-momentum tensor of one of the polarization modes is directly obtained from its effective action as
\begin{align}
T_{\alpha\beta}
&= \frac{G_4(\ph)}{16\pi G}
    \pac{  
        h^\ocircle_{,\alpha} h^\ocircle_{,\beta}
        - \frac{1}{2} (h^\ocircle_{,\gamma} h_\ocircle^{,\gamma})
        \, \eta_{\alpha\beta}
        } \\
&= \chi_{,\alpha} \chi_{,\beta}
    - \frac{1}{2} (\chi_{,\gamma} \chi^{,\gamma})
    \, \eta_{\alpha\beta} \ .
\end{align}
Its quantum version~$\hat{T}_{\alpha\beta}$, which is an operator on a Fock space, is obtained by simply adding hats on $\chi$. The result is then expressed in terms of the creation and annihilation operators as
\begin{multline}
\hat{T}_{\alpha\beta}
= \int \frac{\dd^3\vect{k}}{(2\pi)^3}
        \frac{\dd^3\vect{k}'}{(2\pi)^3}
        \frac{\hbar}{2\sqrt{\omega\omega'}}
        \bigg[
            - \hat{a}_{\vect{k}} \hat{a}_{\vect{k}'}
                \ex{\i(k_\gamma+k'_\gamma) x^\gamma} \\
            + \hat{a}_{\vect{k}} \hat{a}_{\vect{k}'}^\dagger
                \ex{\i(k_\gamma-k'_\gamma) x^\gamma}
            + \hat{a}_{\vect{k}}^\dagger \hat{a}_{\vect{k}'}
                \ex{-\i(k_\gamma-k'_\gamma) x^\gamma} \\
            - \hat{a}_{\vect{k}}^\dagger 
                \hat{a}_{\vect{k}'}^\dagger
                \ex{-\i(k_\gamma+k'_\gamma) x^\gamma}
        \bigg]
        \pa{k_\alpha k'_\beta - \frac{1}{2} k^\gamma k'_\gamma \, \eta_{\alpha\beta}} .
\end{multline}

Let $\Sigma$ be a spatial slice of $\region$ with $t=\cst$. The total energy in $\Sigma$ plays the role of the Hamiltonian, and reads\footnote{This result is only approximate, due to the finiteness of $\region$.}
\begin{equation}
\hat{E} \define \int_\Sigma \dd^3 x \;
        \pa{\hat{T}_{00}
           -\langle0|\hat{T}_{00}|0\rangle}
= \int \frac{\dd^3\vect{k}}{(2\pi)^3} \;
        \hbar\omega \,
        \hat{a}^\dagger_{\vect{k}} \hat{a}_{\vect{k}} ,
\end{equation}
where the energy of vacuum has been subtracted as usual. One-particle states of the form $\ket{\vect{k}}\define \hat{a}^\dagger_{\vect{k}}\ket{0}$ are clearly eigenstates of $\hat{E}$, with eigenvalue $\hbar\omega$. Hence, they qualify as gravitons, according to our definition. It can be noted that they are not eigenstates of the spatial part of the momentum operator
\begin{equation}
\hat{P}^\alpha
\define \int_\Sigma \dd^3 x \;
        \pa{\hat{T}^{0\alpha}
           -\langle0|\hat{T}^{0\alpha}|0\rangle} \ ,
\end{equation}
for which terms of the form $\hat{a}_{\vect{k}}\hat{a}_{-\vect{k}}$ and $\hat{a}^\dagger_{\vect{k}}\hat{a}^\dagger_{-\vect{k}}$ remain. However, the expectation value of $\hat{P}^\alpha$ on a one-particle state reads
\begin{equation}
\langle \vect{k} | \hat{P}^\alpha | \vect{k} \rangle
= \hbar k^\alpha \ ,
\end{equation}
which we shall call a quantum of gravitational energy-momentum.

Summarizing, in any small region~$\region$ of spacetime, the transverse metric perturbation modes~$h_\circlearrowleft$ and $h_\circlearrowright$ essentially behave as canonical scalar fields. Their kinetic coupling with~$G_4(\ph)$ effectively translates as a negligible mass~$m^2=\Box \sqrt{G_4}/\sqrt{G_4}$. Their quantization then leads to very standard results; in particular, a graviton with wave vector~$\vect{k}$ carries an energy~$E\e{g}=\hbar\omega$, and a linear momentum~$\vect{p}\e{g}=\hbar\vect{k}$. Note that $\omega^2=\vect{k}^2+m^2\approx \vect{k}^2$ is observer dependent. Here, it was implicitly defined with respect to a normal coordinate system~$(x^\alpha)$. In general, an observer with four-velocity~$u^\mu$ would measure $\omega=-u_\mu k^\mu$.

\subsection{Graviton-number conservation}

Let us finally show how the classical relation~\eqref{eq:propagation_GW_transverse_amplitude} may be related to the conservation of graviton number. We shall follow the same rationale as Ref.~\cite{Fleury:2015hgz}, Sec.~1.3.2, which concerned photon-number conservation.

Consider an arbitrary observer in spacetime with four-velocity~$u^\mu$. From the expression~\eqref{eq:energy_momentum_tensor_GW_final} of the energy-momentum tensor~$T_{\mu\nu}\h{GW}$ of a GW, we conclude that its four-momentum density with respect to the observer is
\begin{equation}
P^\mu\e{GW}
\define - u_\nu T^{\mu\nu}\e{GW}
= \frac{G_4\omega}{64\pi G} ||H_\perp||^2 k^\mu \ ,
\end{equation}
with $\omega\define -u_\nu k^\nu$. Physically speaking, $\rho\e{GW}\define -u_\mu P^\mu\e{GW}$ is the energy density of the GW in the observer's frame. Besides, the spatial projection $\Pi^\mu\e{GW}\define (u^\mu u_\nu + \delta^\mu_\nu) P\e{GW}^\nu$ on the observer's local space represents the density of 3-momentum of the GW, as well as its energy flux density in the observer's frame.

It is, therefore, very natural to define the graviton flux density four-vector as
\begin{equation}
J^\mu \define \frac{P^\mu\e{GW}}{\hbar\omega}
= \frac{G_4}{64\pi G \hbar} \, ||H_\perp||^2 k^\mu \ .
\end{equation}
Indeed, with such a definition, $n\define -u_\mu J^\mu = \rho\e{GW}/E\e{g}$ clearly represents the number of gravitons per unit volume, while $(u^\mu u_\nu + \delta^\mu_\nu) J^\nu=\Pi^\mu\e{GW}/p\e{g}$ represents the graviton flux density, i.e. the number of gravitons crossing an arbitrary surface, per unit time and per unit area. Note that, contrary to $P^\mu\e{GW}$, $J^\mu$ is observer independent.

The propagation equation \eqref{eq:propagation_GW_transverse_amplitude} implies that $J^\mu$ is divergence free. Indeed, for each polarization amplitude $H_\circlearrowleft, H_\circlearrowright$,
\begin{equation}
    (G_4|H_\ocircle|^2 k^\mu)_{;\mu} =0 \ , 
\end{equation}
so that finally
\begin{empheq}[box=\fbox]{equation}
\label{eq:graviton_conservation}
J\indices{^\mu_;_\mu} = 0 \ .
\end{empheq}

Equation~\eqref{eq:graviton_conservation} must be understood as a continuity equation. Consider indeed an arbitrary observer, and a small spatial domain~$\Sigma$ in their frame. Let $(x^\alpha)=(t,x^i)$ be a normal coordinate system adapted to $\Sigma$, i.e. such that $\bar{g}_{\alpha\beta}\approx\eta_{\alpha\beta}$ across $\Sigma$. Integrating Eq.~\eqref{eq:graviton_conservation} over that spatial domain then leads to
\begin{equation}
\partial_t \int_\Sigma \dd^3 x \; n
= - \int_{\partial\Sigma} \dd A \; \vect{n}\cdot\vect{J} \ , 
\end{equation}
where $\vect{n}$ is the outgoing normal vector orthogonal to the boundary~$\partial\Sigma$ of the spatial domain $\Sigma$. In other words, the variation of the total number of gravitons in $\Sigma$ is given by how many gravitons enter through its boundary.

\section{Summary and conclusion}
\label{sec:conclusion}

In this article, we have investigated the propagation of GWs in Horndeski theories of gravity. Given the strict observational constraints set by GW170817/GRB~170817, we restricted our discussion to the class of models in which GWs propagate at the speed of light. These consist of a combination of k-essence~$G_2(\ph, X)$ conformally coupled spacetime geometry, $G_4(\ph)R$, and a cubic-Galileon kinetic braiding~$G_3(\ph, X)\Box\ph$. Only the last two can potentially affect the propagation of GWs.

We have found that, on the one hand, kinetic braiding introduces a nonphysical longitudinal mode~$h^{||}_{\mu\nu}$ with no observable impact. Conformal coupling, on the other hand, affects the amplitude of GWs without modifying their polarization, which is parallel transported. The main consequence is that the \emph{gravitational distance}~$D\e{G}$ measured with standard sirens reads
\begin{equation}\label{eq:DG_DL_conclusion}
D\e{G} = \sqrt{\frac{G_4(\ph\e{o})}{G_4(\ph\e{s})}} \, D\e{L}
        = \frac{M\e{o}}{M\e{s}} D\e{L} 
\end{equation}
where $D\e{L}$ is the electromagnetic luminosity distance, and $\ph\e{s}, \ph\e{o}$ are the local values of the scalar field at the emission and reception events. Equation~\eqref{eq:DG_DL_conclusion} had already been derived for GWs propagating on a homogeneous-isotropic cosmological background, with a homogeneous scalar field. The increment of the present article is the rigorous proof of the validity of Eq.~\eqref{eq:DG_DL_conclusion} for \emph{any} background spacetime, and for \emph{any} distribution for the scalar field. In particular, it holds if the GW is lensed; or if it is emitted, received, or propagates through screened regions. Furthermore, we have demonstrated that Eq.~\eqref{eq:DG_DL_conclusion} corresponds to the conservation of the number of gravitons--a property that was already suspected in Ref.~\cite{Belgacem:2018lbp}.

Albeit very general, our analysis relies on two important assumptions: (a) scalar waves can be neglected in the harmonic gauge; and (b) GWs are in the geometric-optics regime (eikonal approximation). The latter assumption is easily satisfied as long as spacetime geometry and the scalar field vary over astronomical scales, but it breaks down in the vicinity of compact objects, and for very low-frequency GWs. The former assumption is justified by (i) the suppression of scalar radiation for screened emitters, and (ii) the weak coupling between tensor and scalar perturbations, which limits the conversion of tensor waves into scalar waves. This point may nevertheless deserve further investigation, which was beyond the scope of this article. 

Standard sirens have been argued to be a key probe of variations of the effective Planck mass~$M$ in Horndeski theories. However, since $D\e{G}$ depends on \emph{local} values of the scalar field at emission and reception, screening mechanisms may significantly limit the efficiency of this probe; see Ref.~\cite{Dalang:2019fma} for a more detailed discussion. As Horndeski escapes the sirens, we may recall the words of Athena at the ending of the Odyssey,

\medskip

\begin{flushright}
{\sl
\indent Horndeski, you are adaptable;\\
\indent you always find solutions.
}
\end{flushright}

\appendix
\section{MASSLIKE TERMS IN GW PROPAGATION}
\label{app:massterms}

In Sec.~\ref{subsubsec:kinetic_damping_mass}, we split the linearized equation of motion for $h_{\mu\nu}$ into kinetic, damping, and masslike terms. What we called masslike terms were all the terms of the form~$\mass^i_{\mu\nu}=\mass^i_{\mu\nu\rho\sigma} h^{\rho\sigma}$, where $\mass^i_{\mu\nu\rho\sigma}$ is a tensor depending on background quantities only. Their expressions are
\begin{widetext}
\begin{align}
\label{eq:mass_term_2}
\mass^2_{\mu\nu\rho\sigma}
&= -\frac{1}{4} \pa{
                    G_{2,XX} \ph_{,\mu}\ph_{,\nu} + G_{2,X} \bar{g}_{\mu\nu}
                    }
                \ph_{,\rho} \ph_{,\sigma}
    - \frac{1}{2} G_2 \bar{g}_{\mu\rho} \bar{g}_{\nu\sigma}\,, \\
\label{eq:mass_term_3}
\mass^3_{\mu\nu\rho\sigma}
&= -\frac{1}{2}
    \bigg[
        G_{3,XX} \ph_{,(\mu} X_{,\nu)}
        + (G_{3,\ph X} + \frac{1}{2} G_{3,XX} \Box\ph) \ph_{,\mu}\ph_{,\nu}
        +   \left(
                G_{3,\ph} + X G_{3,\ph X}
                - \frac{1}{2} G_{3,XX} \ph^{,\rho} X_{,\rho}
            \right) \bar{g}_{\mu\nu}
    \bigg] \ph_{,\rho} \ph_{,\sigma} \nonumber\\
& \qquad
    - \left(
        G_{3,\ph} X
        - \frac{1}{2} G_{3,X} \ph^{,\lambda} X_{,\lambda}
    \right) \bar{g}_{\mu\rho} \bar{g}_{\nu\sigma}
- G_{3,X}   \left[
            \ph_{,\rho} \ph_{;\sigma(\mu} \ph_{,\nu)}
            - \frac{1}{2} \ph_{,\mu}\ph_{,\nu} \ph_{;\rho\sigma}
            + \frac{1}{2}
                \left(
                    \ph_{,\rho} X_{,\sigma}
                    - \ph^{,\lambda}\ph_{,\rho}\ph_{;\sigma\lambda}
                \right) \bar{g}_{\mu\nu}
            \right]\,, \\
\label{eq:mass_term_4}
\mass^4_{\mu\nu\rho\sigma}
&= \frac{G_4}{2} \pa{ 2\bar{R}_{\rho\mu\nu\sigma}
                    + 2 \bar{R}_{\rho(\mu} \bar{g}_{\nu)\sigma}
                    + \bar{R}_{\rho \sigma} \bar{g}_{\mu \nu}
                    - \bar{R} \bar{g}_{\mu\rho} \bar{g}_{\nu\sigma}
                    }
    - G_{4;\rho\sigma} \bar{g}_{\mu\nu}
    + \Box G_4 \, \bar{g}_{\mu\rho}\bar{g}_{\nu\sigma} \, .
\end{align}
\end{widetext}
Note that, in Eqs.~\eqref{eq:mass_term_2}, \eqref{eq:mass_term_3}, \eqref{eq:mass_term_4}, $X$ refers to the background $\bar{X}=-\bar{g}^{\mu\nu}\ph_{,\mu}\ph_{,\nu}/2$.

\section{NULL TETRAD}
\label{app:null_tetrad}

A null tetrad may be strictly defined from its algebraic properties, as in Sec.~\ref{subsubsec:tetrad_decomposition}. However, its physical meaning is better understood if one constructs it starting from timelike and spacelike vectors. Let $u^\mu$ be the four-velocity of an arbitrary observer, normalized with respect to the background metric,
\begin{equation}
    \bar{g}_{\mu\nu} u^\mu u^\nu = -1.
\end{equation}
For a wave propagating along $k^\mu$, one defines the unit four-vector~$d^\mu$ associated with the spatial propagation direction of the wave in the observer's frame as
\begin{equation}
    k^\mu = \omega(u^\mu+d^\mu) \ ,
\end{equation}
with $\omega=-\bar{g}_{\mu\nu}u^\mu k^\nu$. The auxiliary null vector~$n^\mu$ can then be defined as, e.g.,
\begin{equation}
n^\mu = \frac{1}{2\omega} \, (d^\mu-u^\mu) \ .
\end{equation}

The piece of plane orthogonal to both $u^\mu, d^\mu$ can be seen as a spatial screen, orthogonal to the wave's propagation direction. Introduce an orthonormal basis $(s_1^\mu, s_2^\mu)$ of that screen, i.e.
\begin{gather}
\bar{g}_{\mu\nu} u^\mu s_A^\nu
= \bar{g}_{\mu\nu} d^\mu s_A^\nu
= 0 \,,\\
\bar{g}_{\mu\nu} s_A^\mu s_B^\nu = \delta_{AB} \ ,
\end{gather}
where indices $A, B$ take values in $\{1,2\}$. The last piece of the null tetrad can then be defined as
\begin{equation}
m^\mu = \frac{1}{\sqrt{2}} (s_1^\mu + \i s_2^\mu) \ .
\end{equation}

The set $(u^\mu, d^\mu, s_1^\mu, s_2^\mu)$ forms a "traditional" tetrad, with one timelike and three spacelike vectors. The set $[k^\mu, n^\mu, m^\mu, (m^\mu)^*]$, besides, forms a null tetrad with the desired properties; all its vectors are null, and the only nonvanishing scalar products are $n_\mu k^\mu=m^\mu m_\mu^*=1$. The background metric is expressed in terms of both tetrads as
\begin{align}
\bar{g}_{\mu\nu}
&= - u_\mu u_\nu + d_\mu d_\nu
    + \delta_{AB} s^A_\mu s^B_\nu \\
&= 2k_{(\mu} n_{\nu)} + 2m_{(\mu} m^*_{\nu)} \ .
\end{align}

The transverse-traceless amplitude~$H_{\mu\nu}^\perp$ of a GW can be decomposed either in terms of the traditional tetrad, which defines the usual plus ($+$) and cross ($\times$) linear polarizations; or in terms of the null tetrad, which defines the left ($\circlearrowleft$) and right ($\circlearrowright$) circular polarizations,
\begin{align}
H_{\mu\nu}^\perp
&= H_+ (s^1_\mu s^1_\nu - s^2_\mu s^2_\nu)
    + H_\times (s^1_\mu s^2_\nu + s^2_\mu s^1_\nu) \\
&= H_\circlearrowleft m_\mu m_\nu
    + H_\circlearrowright m^*_\mu m^*_\nu \ .
\end{align}
These are, therefore, related by
\begin{align}
\label{eq:left_plus_cross}
H_\circlearrowleft &= H_+ - \i H_\times \ ,
\\
\label{eq:right_plus_cross}
H_\circlearrowright &= H_+ + \i H_\times \ .
\end{align}

\section{COMPLEX INNER PRODUCT}
\label{app:null_product}

Working with a null basis can be very counterintuitive, especially if one wants to extract the components of a tensor over that basis. As a simple illustration, if a vector $v_\mu$ is decomposed over the null tetrad as
\begin{equation}
v_\mu
= v_k k_\mu + v_n n_\mu + v_m m_\mu + v_{m^*} m_\mu^* \ ,
\end{equation}
then, e.g., $v_k \not= v^\mu k_\mu = v_n$ contrary to what is expected from a natural Riemannian (rather than Lorentzian) intuition.

That intuition can nevertheless be restored by introducing an inner product better adapted to that purpose. Let us define the dual of $v_\mu$ as
\begin{equation}
\star v_\mu =
v_k^* n_\mu + v_n^* k_\mu + v_m^* m^*_\mu + v_{m^*}^* m_\mu
\ ,
\end{equation}
where $*$ denotes the normal complex conjugation; in other words, $\star$ takes the complex conjugate of the components of $v_\mu$ over the null basis, and transforms the basis vectors as
\begin{equation}
\star k_\mu=n_\mu \quad
\star n_\mu=k_\mu \quad
\star m_\mu=m_\mu^* \quad
\star m_\mu^*=m_\mu \ .
\end{equation}

The $\star$ duality is clearly an involution, $(\star)^2=1$, and defines the complex inner product $u^\mu\star v_\mu$ for any two four-vectors $u_\mu, v_\mu$. This product is Hermitian since $v^\mu\star u_\mu=(u^\mu\star v_\mu)^*$. It is also positive definite since the associated norm reads
\begin{equation}
||v||^2 \define
v^\mu \star v_\mu
= |v_k|^2 + |v_n|^2 + |v_m|^2 + |v_{m^*}|^2 \ .
\end{equation}

Importantly, in terms of the $\star$ inner product, the null tetrad $(e_\mu^\alpha)=(k_\mu, n_\mu, m_\mu, m^*_\mu)$ is \emph{orthonormal} in a Riemannian sense,
\begin{equation}
e_\mu^\alpha \star e^\mu_\beta = \delta^\alpha_\beta \ .
\end{equation}
This implies that the components of a vector are directly extracted by taking the inner product with the associated basis vector, $v_\alpha = v_\mu \star e^\mu_\alpha$.

The above notions are then immediately extended to tensors of any rank with the rule $\star(u_\mu v_\nu)=(\star u_\mu)(\star v_\nu)$. For example, the following rank-two tensors introduced in Sec.~\ref{subsubsec:parallel_transport},
\begin{align}
e_{\mu\nu}^\circlearrowleft
&\define m_\mu m_\nu
\\
e_{\mu\nu}^\circlearrowright
&\define m^*_\mu m^*_\nu \ ,
\end{align}
form an orthonormal set:
\begin{align}
e_{\mu\nu}^\circlearrowleft\star e^{\mu\nu}_\circlearrowleft
&= m^\mu m^\nu m_\mu^* m_\nu^* = 1\,,
\\
e_{\mu\nu}^\circlearrowright\star e^{\mu\nu}_\circlearrowright
&= m^*_\mu m^*_\nu m^\mu m^\nu = 1 \,,
\\
e_{\mu\nu}^\circlearrowleft\star e^{\mu\nu}_\circlearrowright
&= m_\mu m_\nu m^\mu m^\nu = 0 \ .
\end{align}

Similarly, the rank-four tensors
\begin{align}
e^\circlearrowleft_{\mu\nu\rho\sigma}
&= 2 m_{[\mu} k_{\nu]} m_{[\rho} k_{\sigma]}\,,
\\
e^\circlearrowright_{\mu\nu\rho\sigma}
&= 2 m^*_{[\mu} k_{\nu]} m^*_{[\rho} k_{\sigma]}\,,
\end{align}
form an orthonormal set
\begin{align}
e_{\mu\nu\rho\sigma}^\circlearrowleft\star e^{\mu\nu\rho\sigma}_\circlearrowleft
&= 4 m^{[\mu} k^{\nu]} m^{[\rho} k^{\sigma]}
    m^*_{[\mu} n_{\nu]} m^*_{[\rho} n_{\sigma]}
= 1\, ,
\\
e_{\mu\nu\rho\sigma}^\circlearrowright\star e^{\mu\nu\rho\sigma}_\circlearrowright
&= 4 m^*_{[\mu} k_{\nu]} m^*_{[\rho} k_{\sigma]}
    m^{[\mu} n^{\nu]} m^{[\rho} n^{\sigma]}
= 1\, ,
\\
e_{\mu\nu\rho\sigma}^\circlearrowleft\star e^{\mu\nu\rho\sigma}_\circlearrowright
&= 4 m^{[\mu} k^{\nu]} m^{[\rho} k^{\sigma]}
     m_{[\mu} n_{\nu]} m_{[\rho} n_{\sigma]}
= 0\, .
\end{align}

\section{REMOVING THE LONGITUDINAL MODE?}
\label{app:gauge_removal_longitudinal}

We have seen in Sec.~\ref{subsec:longitudinal_transverse_modes} that, in the harmonic gauge, the GW amplitude can be decomposed as the sum of a longitudinal and a transverse mode, $H_{\mu\nu} = H^{||}_{\mu\nu} + H^\perp_{\mu\nu}$, where
\begin{equation}
\label{eq:longitudinal_mode}
H^{||}_{\mu\nu}=2 H_{(\mu} k_{\nu)}
\end{equation}
involves a vector field~$H_\mu$ whose equation of motion follows from Eq.~\eqref{eq:propagation_GW_amplitude}. Besides, in the geometric optics regime, we have seen that the longitudinal mode produces no tidal force (\ref{subsubsec:longitudinal_nonphysical}) and carries no energy-momentum (\ref{subsec:energy-momentum_GW}). This mode being nonphysical, we naturally expected it to be removable by a suitable gauge transformation.

Let $\xi_\mu$ be a gauge field generating the gauge transformation~$h_{\mu\nu}\mapsto h_{\mu\nu}+\xi_{(\mu;\nu)}$. In the geometric optics regime, it is convenient to write the gauge field as a wave
\begin{equation}
\xi_\mu = \frac{1}{2}\Xi_\mu \ex{\i w} + \cc \ ,
\end{equation}
where $w$ is the same phase as the GW's, and $\Xi_\mu$ is a complex amplitude. Indeed, with this ansatz, the gauge transformation of the GW amplitude simply reads
\begin{equation}
H_{\mu\nu} \mapsto H_{\mu\nu} + 2\Xi_{(\mu} k_{\nu)} \ .
\end{equation}

Comparing with Eq.~\eqref{eq:longitudinal_mode}, one may immediately conclude that choosing $\Xi_\mu=-H_\mu$ would remove the longitudinal mode. In GR, this operation can be performed along the whole geodesic connecting the source to the observer. Here, however, it can only be done locally. This is because $\Xi_\mu$ and $H_\mu$ do not satisfy the same propagation equation if $G_{4,\ph}\not= 0$ or $G_{3,X}\not = 0$. Let us elaborate on that point. For the gauge field to preserve the harmonic gauge condition $\gamma\indices{_\mu_\nu^;^\nu}=0$, one must ensure that $\Box\xi_\mu + \bar{R}^\nu_\mu \xi_\nu \approx \Box\xi_\mu = 0$; that is, in the geometric optics regime
\begin{align}
k^\mu k_\mu &= 0 \ , \\
\propagator \Xi_\mu &= 0 \ ,
\end{align}
with $k_\mu\define w_{,\mu}$, and $\propagator = 2k^\mu\bar{\nabla}_\mu+k\indices{^\mu_;_\mu}$ as in Eq.~\eqref{eq:definition_propagator}. The first condition is trivially satisfied, but the second one turns out to be incompatible with the propagation equation for $H_\mu$. Indeed, contracting Eq.~\eqref{eq:propagation_GW_amplitude} with $n^\mu (\delta^\nu_\rho - \frac{1}{2} n^\nu k_\rho)$, and renaming indices, we find
\begin{equation}\label{eq:Longitudinal_Amplitude_Propagation}
G_4 \propagator H_\mu
= G_4^{,\nu} (H^\perp_{\mu\nu} + H_\nu k_\mu)
    - \frac{1}{2} G_{3,X}
        \ph^{,\rho} \ph^{,\sigma} H_{\rho\sigma}
        \ph_{,\mu} \ .
\end{equation}
Thus, one cannot gauge the longitudinal mode away along the whole GW's worldline without violating the harmonic gauge condition, which was necessary for all the above. Note, however, that we may still set $H_\mu=0$ at any point, consistently with the nonphysical character of the longitudinal mode.

The reason why we cannot fully eliminate $H^{||}_{\mu\nu}$, despite its absence of physical content, is not entirely clear. The most plausible explanation is that it is a consequence of neglecting scalar waves. Alternatively, $H^{||}_{\mu\nu}$ may possess physical information beyond geometric optics.

\section{COMPARISON WITH ANOTHER RECENT WORK}
\label{app:comparison_G19}

Shortly after the first preprint version of our article came out, an independent work by Garoffolo \textit{et al}.~\cite{Garoffolo:2019mna}, hereafter G19, appeared on the same topic. Since our respective approaches and conclusions are slightly different, we shall compare them explicitly in this appendix\footnote{The comparison will be based on the second version (\href{https://arxiv.org/abs/1912.08093v2}{v2}) of their preprint.}.

\subsection{Scalar waves and gauge fixing}

A first difference between G19 and our work is the way to deal with perturbations of the scalar field. While we choose to neglect scalar perturbations in the harmonic gauge (see Sec.~\ref{subsec:scalar_waves}), G19 proposes to eliminate them with a more elaborate gauge choice. Specifically, G19 imposes at the same time the harmonic gauge, $\gamma\indices{_\mu_\nu^;^\nu}=0$ and a kind of unitary gauge for the wave-part of the scalar perturbation, $\delta\ph=0$; let us call \emph{harmonic-unitary gauge} the resulting setup. As demonstrated in G19's section~2.5, a \emph{necessary} condition to enforce the harmonic-unitary gauge is that scalar and tensor waves propagate at the same speed, i.e. here $\Box\delta\ph=0$ at the relevant order of approximation. This condition is quite restrictive. In the framework of luminal Horndeski theories, and excluding fine-tuning, G19's Appendix~A shows that $\Box\delta\ph=0$ holds only if $G_{2,XX}=G_{3,X}=0$. Since, in that case,
\begin{equation}
\Lagrangian_3
= G_3(\ph)\Box\ph
= [G_3(\ph)\,\ph^{,\mu}]_{;\mu} + 2X G_{3,\ph}(\ph) \,,
\end{equation}
belongs to the $\Lagrangian_2$ class up to a total derivative, the potential applicability of the harmonic-unitary gauge is thereby restricted to conformally coupled quintessence.

Besides its restricted range of applicability, one may note that $\Box\delta\ph=0$ is a necessary but not sufficient condition to enforce the harmonic-unitary gauge. A simple counterexample is minimally coupled quintessence, $\Lagrangian_2=X, \Lagrangian_4=R$, in vacuum. In that case $\bar{\ph}=\cst$, so that any scalar perturbation $\delta\ph$ is gauge invariant, and hence cannot be gauged away. One may object that in reality $\bar{\ph}$ always exhibits some gradient, but even in that case counterexamples can be found.

Since the applicability of the harmonic-unitary gauge is quite uncertain, we conclude that G19's subsequent analysis of GWs in luminal Horndeski theories (Sec.~4 of G19) must generally make the same assumption as we did, namely that scalar waves are negligible in the harmonic gauge. This is especially true when kinetic braiding terms, proportional to $G_{3,X}$, are explicitly kept in the GW propagation equations.

\subsection{Parallel transport of the GW polarization}

The second important discrepancy between G19 and the present work concerns the transport of the GW polarization. While we find in Sec.~\ref{subsubsec:parallel_transport} that the polarization is parallel transported, G19 finds that it is not in general [see e.g. Eqs.~(5.2), (5.73) therein]. This discrepancy is due to different definitions of the polarization tensor.

Specifically, we choose to define the polarization tensor from the observable transverse mode~$H_{\mu\nu}^\perp$, while G19 defines it from the amplitude $\Gamma_{\mu\nu}$ of the trace-reversed metric perturbation~$\gamma_{\mu\nu}$,
\begin{align}
\label{eq:polarization_this_work}
\text{[this work]} \qquad
e_{\mu\nu}
&\define \frac{H^\perp_{\mu\nu}}{||H_\perp||} \ ,
\\
\label{eq:polarization_G19}
\text{[G19]}
\qquad
e_{\mu\nu}\h{G19}
&\define \frac{\Gamma_{\mu\nu}}
    {\sqrt{\Gamma^{\mu\nu}\Gamma^*_{\mu\nu}}} \ .
\end{align}
With the definition\footnote{We added a complex conjugation in the normalization of $e_{\mu\nu}\h{G19}$ with respect to G19, in order to allow for complex amplitudes. This does not change the results of G19's calculations.}~\eqref{eq:polarization_G19}, G19 finds
\begin{multline}
\label{eq:departure_parallel_transport_G19}
k^\rho e\h{G19}_{\mu\nu;\rho}
=
\frac{G_{4,\ph}}{G_4} \,
    \ph^{,\rho} e\h{G19}_{\rho(\mu} k_{\nu)} \\
- \frac{1}{2} \frac{G_{3,X}}{G_4}
    \ph^{,\rho} \ph^{,\sigma} e\h{G19}_{\rho\sigma}
    \l[ 2 k_{(\mu} \ph_{,\nu)} 
        - \frac{1}{2}\bar{g}_{\mu\nu}
            k^\tau \ph_{,\tau}
    \r] ,
\end{multline}
which is nonzero in general.

There are three differences between Eq.~\eqref{eq:polarization_G19} and our definition~\eqref{eq:polarization_this_work}: (i) the use of the trace-reversed metric perturbation; (ii) the presence of the longitudinal mode; and (iii) a different normalization. Nevertheless, the departure from parallel transport in $e_{\mu\nu}\h{G19}$ somehow boils down to the nontrivial propagation of the longitudinal mode. This may be understood by writing explicitly $\Gamma_{\mu\nu}$. Using the decomposition~\eqref{eq:longitudinal_and_transverse} of $H_{\mu\nu}$ into longitudinal and transverse modes, we find
\begin{align}
\Gamma_{\mu\nu}
\define H_{\mu\nu} - \frac{1}{2} H \bar{g}_{\mu\nu}
= \Gamma_{\mu\nu}^{\perp} +  \Gamma_{\mu\nu}^{||} \ ,
\end{align}
with
\begin{align}
\label{eq:Gamma_perp}
\Gamma^\perp_{\mu\nu}
&\define H_{\mu\nu}^\perp - 2H_n m_{(\mu}m^*_{\nu)} \\
\Gamma_{\mu\nu}^{||}
&= 2\left[
        H_k k_{(\mu} + H_m m_{(\mu} + H_{m^*} m^*_{(\mu}
        \right] k_{\nu)} \ ,
\end{align}
and where $H_{\mu}$ is the vector field involved in the longitudinal mode, $H^{||}_{\mu\nu}=2H_{(\mu}k_{\nu)}$. The normalization of $e_{\mu\nu}\h{G19}$ then reads
\begin{equation}
\Gamma^{\mu\nu}\Gamma^*_{\mu\nu}
=
\Gamma_\perp^{\mu\nu}(\Gamma_\perp^*)_{\mu\nu}
= ||H_\perp||^2 + 2|H_n|^2 \ .
\end{equation}

G19 did not consider $\Gamma^{||}_{\mu\nu}$ in their decomposition~(3.24) for simplicity, which does not affect the normalization $\Gamma^{\mu\nu}\Gamma^*_{\mu\nu}$, although the other consequences of that assumption are unclear. Besides, while G19 interprets the second term~$2H_n m_{(\mu}m^*_{\nu)}$ of Eq.~\eqref{eq:Gamma_perp} as a scalar mode, it is worth emphasizing that this mode is nonphysical--just like $H_{\mu}$ is not a physical longitudinal mode. Specifically, a physical scalar mode would have the same form $\propto m_{(\mu}m^*_{\nu)}$, but in $H_{\mu\nu}$ rather than in $\Gamma_{\mu\nu}$. Such a mode would be associated with Ricci waves, and would thereby produce isotropic oscillating tidal forces in screen space.

Summarizing, all the differences between $e_{\mu\nu}$ and $e\h{G19}_{\mu\nu}$ are encoded in the vector field $H_\mu$. The right-hand side of Eq.~\eqref{eq:departure_parallel_transport_G19} then originates from the action of a covariant derivative on $H_\mu$ in the direction of $k^\mu$. This can be checked using Eqs.~\eqref{eq:propagation_GW_transverse_amplitude}, \eqref{eq:Longitudinal_Amplitude_Propagation}.


We stress that our choice~$e_{\mu\nu}$ for the polarization tensor, based on the sole transverse mode $H^\perp_{\mu\nu}$, is observationally justified because it is directly related to curvature perturbations, as shown in Sec.~\ref{subsubsec:parallel_transport}. The use of the trace-reversed metric perturbation, or the inclusion of the nonphysical longitudinal mode, do not share this motivation. Importantly, they lead to spurious departures from parallel transport, just like the polarization of light, if defined from $A_\mu$, is not parallel transported if one includes nonphysical gauge modes.

\acknowledgements

We thank Homer for inspiration and Gregory Horndeski for kindly accepting the role of Odysseus. We also thank Giulia Cusin and Karim Noui for useful discussions, and Ryan McManus for providing references about scalar radiation. C.D. and L.L. were supported by a Swiss National Science Foundation (SNSF) Professorship grant (No.~170547). P.F. acknowledges the financial support of the Swiss National Science Foundation during the first part of this project. The project that gave rise to these results received the support of a fellowship from ``la Caixa'' Foundation (ID 100010434). The fellowship code is LCF/BQ/PI19/11690018.

\bibliography{bibliography_Horndeski_GW.bib}

\begin{thebibliography}{86}%
\makeatletter
\providecommand \@ifxundefined [1]{%
 \@ifx{#1\undefined}
}%
\providecommand \@ifnum [1]{%
 \ifnum #1\expandafter \@firstoftwo
 \else \expandafter \@secondoftwo
 \fi
}%
\providecommand \@ifx [1]{%
 \ifx #1\expandafter \@firstoftwo
 \else \expandafter \@secondoftwo
 \fi
}%
\providecommand \natexlab [1]{#1}%
\providecommand \enquote  [1]{``#1''}%
\providecommand \bibnamefont  [1]{#1}%
\providecommand \bibfnamefont [1]{#1}%
\providecommand \citenamefont [1]{#1}%
\providecommand \href@noop [0]{\@secondoftwo}%
\providecommand \href [0]{\begingroup \@sanitize@url \@href}%
\providecommand \@href[1]{\@@startlink{#1}\@@href}%
\providecommand \@@href[1]{\endgroup#1\@@endlink}%
\providecommand \@sanitize@url [0]{\catcode `\\12\catcode `\$12\catcode
  `\&12\catcode `\#12\catcode `\^12\catcode `\_12\catcode `\%12\relax}%
\providecommand \@@startlink[1]{}%
\providecommand \@@endlink[0]{}%
\providecommand \url  [0]{\begingroup\@sanitize@url \@url }%
\providecommand \@url [1]{\endgroup\@href {#1}{\urlprefix }}%
\providecommand \urlprefix  [0]{URL }%
\providecommand \Eprint [0]{\href }%
\providecommand \doibase [0]{http://dx.doi.org/}%
\providecommand \selectlanguage [0]{\@gobble}%
\providecommand \bibinfo  [0]{\@secondoftwo}%
\providecommand \bibfield  [0]{\@secondoftwo}%
\providecommand \translation [1]{[#1]}%
\providecommand \BibitemOpen [0]{}%
\providecommand \bibitemStop [0]{}%
\providecommand \bibitemNoStop [0]{.\EOS\space}%
\providecommand \EOS [0]{\spacefactor3000\relax}%
\providecommand \BibitemShut  [1]{\csname bibitem#1\endcsname}%
\let\auto@bib@innerbib\@empty
\bibitem [{\citenamefont {Homer}(2017)}]{Odyssey}%
  \BibitemOpen
  \bibfield  {author} {\bibinfo {author} {\bibnamefont {Homer}},\ }\href@noop
  {} {\emph {\bibinfo {title} {The Odyssey}}}\ (\bibinfo  {publisher} {W. W.
  Norton \& Company},\ \bibinfo {year} {2017})\BibitemShut {NoStop}%
\bibitem [{\citenamefont {Horndeski}(1974)}]{Horndeski:1974wa}%
  \BibitemOpen
  \bibfield  {author} {\bibinfo {author} {\bibfnamefont {G.~W.}\ \bibnamefont
  {Horndeski}},\ }\href {\doibase 10.1007/BF01807638} {\bibfield  {journal}
  {\bibinfo  {journal} {Int. J. Theor. Phys.}\ }\textbf {\bibinfo {volume}
  {10}},\ \bibinfo {pages} {363} (\bibinfo {year} {1974})}\BibitemShut
  {NoStop}%
\bibitem [{\citenamefont {Dvali}\ \emph {et~al.}(2000)\citenamefont {Dvali},
  \citenamefont {Gabadadze},\ and\ \citenamefont {Porrati}}]{Dvali:2000hr}%
  \BibitemOpen
  \bibfield  {author} {\bibinfo {author} {\bibfnamefont {G.~R.}\ \bibnamefont
  {Dvali}}, \bibinfo {author} {\bibfnamefont {G.}~\bibnamefont {Gabadadze}}, \
  and\ \bibinfo {author} {\bibfnamefont {M.}~\bibnamefont {Porrati}},\ }\href
  {\doibase 10.1016/S0370-2693(00)00669-9} {\bibfield  {journal} {\bibinfo
  {journal} {Phys. Lett.}\ }\textbf {\bibinfo {volume} {B485}},\ \bibinfo
  {pages} {208} (\bibinfo {year} {2000})},\ \Eprint
  {http://arxiv.org/abs/hep-th/0005016} {arXiv:hep-th/0005016 [hep-th]}
  \BibitemShut {NoStop}%
\bibitem [{\citenamefont {Nicolis}\ \emph {et~al.}(2009)\citenamefont
  {Nicolis}, \citenamefont {Rattazzi},\ and\ \citenamefont
  {Trincherini}}]{Nicolis:2008in}%
  \BibitemOpen
  \bibfield  {author} {\bibinfo {author} {\bibfnamefont {A.}~\bibnamefont
  {Nicolis}}, \bibinfo {author} {\bibfnamefont {R.}~\bibnamefont {Rattazzi}}, \
  and\ \bibinfo {author} {\bibfnamefont {E.}~\bibnamefont {Trincherini}},\
  }\href {\doibase 10.1103/PhysRevD.79.064036} {\bibfield  {journal} {\bibinfo
  {journal} {Phys. Rev.}\ }\textbf {\bibinfo {volume} {D79}},\ \bibinfo {pages}
  {064036} (\bibinfo {year} {2009})},\ \Eprint {http://arxiv.org/abs/0811.2197}
  {arXiv:0811.2197 [hep-th]} \BibitemShut {NoStop}%
\bibitem [{\citenamefont {Chow}\ and\ \citenamefont
  {Khoury}(2009)}]{Chow:2009fm}%
  \BibitemOpen
  \bibfield  {author} {\bibinfo {author} {\bibfnamefont {N.}~\bibnamefont
  {Chow}}\ and\ \bibinfo {author} {\bibfnamefont {J.}~\bibnamefont {Khoury}},\
  }\href {\doibase 10.1103/PhysRevD.80.024037} {\bibfield  {journal} {\bibinfo
  {journal} {Phys. Rev.}\ }\textbf {\bibinfo {volume} {D80}},\ \bibinfo {pages}
  {024037} (\bibinfo {year} {2009})},\ \Eprint {http://arxiv.org/abs/0905.1325}
  {arXiv:0905.1325 [hep-th]} \BibitemShut {NoStop}%
\bibitem [{\citenamefont {Agarwal}\ \emph {et~al.}(2010)\citenamefont
  {Agarwal}, \citenamefont {Bean}, \citenamefont {Khoury},\ and\ \citenamefont
  {Trodden}}]{Agarwal:2009gy}%
  \BibitemOpen
  \bibfield  {author} {\bibinfo {author} {\bibfnamefont {N.}~\bibnamefont
  {Agarwal}}, \bibinfo {author} {\bibfnamefont {R.}~\bibnamefont {Bean}},
  \bibinfo {author} {\bibfnamefont {J.}~\bibnamefont {Khoury}}, \ and\ \bibinfo
  {author} {\bibfnamefont {M.}~\bibnamefont {Trodden}},\ }\href {\doibase
  10.1103/PhysRevD.81.084020} {\bibfield  {journal} {\bibinfo  {journal} {Phys.
  Rev.}\ }\textbf {\bibinfo {volume} {D81}},\ \bibinfo {pages} {084020}
  (\bibinfo {year} {2010})},\ \Eprint {http://arxiv.org/abs/0912.3798}
  {arXiv:0912.3798 [hep-th]} \BibitemShut {NoStop}%
\bibitem [{\citenamefont {de~Rham}\ and\ \citenamefont
  {Heisenberg}(2011)}]{deRham:2011by}%
  \BibitemOpen
  \bibfield  {author} {\bibinfo {author} {\bibfnamefont {C.}~\bibnamefont
  {de~Rham}}\ and\ \bibinfo {author} {\bibfnamefont {L.}~\bibnamefont
  {Heisenberg}},\ }\href {\doibase 10.1103/PhysRevD.84.043503} {\bibfield
  {journal} {\bibinfo  {journal} {Phys. Rev.}\ }\textbf {\bibinfo {volume}
  {D84}},\ \bibinfo {pages} {043503} (\bibinfo {year} {2011})},\ \Eprint
  {http://arxiv.org/abs/1106.3312} {arXiv:1106.3312 [hep-th]} \BibitemShut
  {NoStop}%
\bibitem [{\citenamefont {Will}(2014)}]{Will:2014kxa}%
  \BibitemOpen
  \bibfield  {author} {\bibinfo {author} {\bibfnamefont {C.~M.}\ \bibnamefont
  {Will}},\ }\href {\doibase 10.12942/lrr-2014-4} {\bibfield  {journal}
  {\bibinfo  {journal} {Living Rev. Rel.}\ }\textbf {\bibinfo {volume} {17}},\
  \bibinfo {pages} {4} (\bibinfo {year} {2014})},\ \Eprint
  {http://arxiv.org/abs/1403.7377} {arXiv:1403.7377 [gr-qc]} \BibitemShut
  {NoStop}%
\bibitem [{\citenamefont {Khoury}\ and\ \citenamefont
  {Weltman}(2004)}]{Khoury:2003aq}%
  \BibitemOpen
  \bibfield  {author} {\bibinfo {author} {\bibfnamefont {J.}~\bibnamefont
  {Khoury}}\ and\ \bibinfo {author} {\bibfnamefont {A.}~\bibnamefont
  {Weltman}},\ }\href {\doibase 10.1103/PhysRevLett.93.171104} {\bibfield
  {journal} {\bibinfo  {journal} {Phys. Rev. Lett.}\ }\textbf {\bibinfo
  {volume} {93}},\ \bibinfo {pages} {171104} (\bibinfo {year} {2004})},\
  \Eprint {http://arxiv.org/abs/astro-ph/0309300} {arXiv:astro-ph/0309300
  [astro-ph]} \BibitemShut {NoStop}%
\bibitem [{\citenamefont {Babichev}\ \emph {et~al.}(2009)\citenamefont
  {Babichev}, \citenamefont {Deffayet},\ and\ \citenamefont
  {Ziour}}]{Babichev:2009ee}%
  \BibitemOpen
  \bibfield  {author} {\bibinfo {author} {\bibfnamefont {E.}~\bibnamefont
  {Babichev}}, \bibinfo {author} {\bibfnamefont {C.}~\bibnamefont {Deffayet}},
  \ and\ \bibinfo {author} {\bibfnamefont {R.}~\bibnamefont {Ziour}},\ }\href
  {\doibase 10.1142/S0218271809016107} {\bibfield  {journal} {\bibinfo
  {journal} {Int. J. Mod. Phys.}\ }\textbf {\bibinfo {volume} {D18}},\ \bibinfo
  {pages} {2147} (\bibinfo {year} {2009})},\ \Eprint
  {http://arxiv.org/abs/0905.2943} {arXiv:0905.2943 [hep-th]} \BibitemShut
  {NoStop}%
\bibitem [{\citenamefont {Vainshtein}(1972)}]{Vainshtein:1972sx}%
  \BibitemOpen
  \bibfield  {author} {\bibinfo {author} {\bibfnamefont {A.~I.}\ \bibnamefont
  {Vainshtein}},\ }\href {\doibase 10.1016/0370-2693(72)90147-5} {\bibfield
  {journal} {\bibinfo  {journal} {Phys. Lett.}\ }\textbf {\bibinfo {volume}
  {39B}},\ \bibinfo {pages} {393} (\bibinfo {year} {1972})}\BibitemShut
  {NoStop}%
\bibitem [{\citenamefont {Babichev}\ \emph {et~al.}(2011)\citenamefont
  {Babichev}, \citenamefont {Deffayet},\ and\ \citenamefont
  {Esposito-Far{\`e}se}}]{Babichev:2011iz}%
  \BibitemOpen
  \bibfield  {author} {\bibinfo {author} {\bibfnamefont {E.}~\bibnamefont
  {Babichev}}, \bibinfo {author} {\bibfnamefont {C.}~\bibnamefont {Deffayet}},
  \ and\ \bibinfo {author} {\bibfnamefont {G.}~\bibnamefont
  {Esposito-Far{\`e}se}},\ }\href {\doibase 10.1103/PhysRevLett.107.251102}
  {\bibfield  {journal} {\bibinfo  {journal} {Phys. Rev. Lett.}\ }\textbf
  {\bibinfo {volume} {107}},\ \bibinfo {pages} {251102} (\bibinfo {year}
  {2011})},\ \Eprint {http://arxiv.org/abs/1107.1569} {arXiv:1107.1569 [gr-qc]}
  \BibitemShut {NoStop}%
\bibitem [{\citenamefont {Wang}\ \emph {et~al.}(2012)\citenamefont {Wang},
  \citenamefont {Hui},\ and\ \citenamefont {Khoury}}]{Wang:2012kj}%
  \BibitemOpen
  \bibfield  {author} {\bibinfo {author} {\bibfnamefont {J.}~\bibnamefont
  {Wang}}, \bibinfo {author} {\bibfnamefont {L.}~\bibnamefont {Hui}}, \ and\
  \bibinfo {author} {\bibfnamefont {J.}~\bibnamefont {Khoury}},\ }\href
  {\doibase 10.1103/PhysRevLett.109.241301} {\bibfield  {journal} {\bibinfo
  {journal} {Phys. Rev. Lett.}\ }\textbf {\bibinfo {volume} {109}},\ \bibinfo
  {pages} {241301} (\bibinfo {year} {2012})},\ \Eprint
  {http://arxiv.org/abs/1208.4612} {arXiv:1208.4612 [astro-ph.CO]} \BibitemShut
  {NoStop}%
\bibitem [{\citenamefont {Beltr{\'a}n~Jim{\'e}nez}\ \emph
  {et~al.}(2016)\citenamefont {Beltr{\'a}n~Jim{\'e}nez}, \citenamefont
  {Piazza},\ and\ \citenamefont {Velten}}]{Jimenez:2015bwa}%
  \BibitemOpen
  \bibfield  {author} {\bibinfo {author} {\bibfnamefont {J.}~\bibnamefont
  {Beltr{\'a}n~Jim{\'e}nez}}, \bibinfo {author} {\bibfnamefont
  {F.}~\bibnamefont {Piazza}}, \ and\ \bibinfo {author} {\bibfnamefont
  {H.}~\bibnamefont {Velten}},\ }\href {\doibase
  10.1103/PhysRevLett.116.061101} {\bibfield  {journal} {\bibinfo  {journal}
  {Phys. Rev. Lett.}\ }\textbf {\bibinfo {volume} {116}},\ \bibinfo {pages}
  {061101} (\bibinfo {year} {2016})},\ \Eprint
  {http://arxiv.org/abs/1507.05047} {arXiv:1507.05047 [gr-qc]} \BibitemShut
  {NoStop}%
\bibitem [{\citenamefont {Creminelli}\ \emph {et~al.}(2020)\citenamefont
  {Creminelli}, \citenamefont {Tambalo}, \citenamefont {Vernizzi},\ and\
  \citenamefont {Yingcharoenrat}}]{Creminelli_2020}%
  \BibitemOpen
  \bibfield  {author} {\bibinfo {author} {\bibfnamefont {P.}~\bibnamefont
  {Creminelli}}, \bibinfo {author} {\bibfnamefont {G.}~\bibnamefont {Tambalo}},
  \bibinfo {author} {\bibfnamefont {F.}~\bibnamefont {Vernizzi}}, \ and\
  \bibinfo {author} {\bibfnamefont {V.}~\bibnamefont {Yingcharoenrat}},\ }\href
  {\doibase 10.1088/1475-7516/2020/05/002} {\bibfield  {journal} {\bibinfo
  {journal} {Journal of Cosmology and Astroparticle Physics}\ }\textbf
  {\bibinfo {volume} {2020}},\ \bibinfo {pages} {002} (\bibinfo {year}
  {2020})}\BibitemShut {NoStop}%
\bibitem [{\citenamefont {Baker}\ \emph {et~al.}(2019)\citenamefont {Baker}
  \emph {et~al.}}]{Baker:2019gxo}%
  \BibitemOpen
  \bibfield  {author} {\bibinfo {author} {\bibfnamefont {T.}~\bibnamefont
  {Baker}} \emph {et~al.},\ }\href@noop {} {\  (\bibinfo {year} {2019})},\
  \Eprint {http://arxiv.org/abs/1908.03430} {arXiv:1908.03430 [astro-ph.CO]}
  \BibitemShut {NoStop}%
\bibitem [{\citenamefont {Gleyzes}\ \emph {et~al.}(2016)\citenamefont
  {Gleyzes}, \citenamefont {Langlois}, \citenamefont {Mancarella},\ and\
  \citenamefont {Vernizzi}}]{Gleyzes:2015rua}%
  \BibitemOpen
  \bibfield  {author} {\bibinfo {author} {\bibfnamefont {J.}~\bibnamefont
  {Gleyzes}}, \bibinfo {author} {\bibfnamefont {D.}~\bibnamefont {Langlois}},
  \bibinfo {author} {\bibfnamefont {M.}~\bibnamefont {Mancarella}}, \ and\
  \bibinfo {author} {\bibfnamefont {F.}~\bibnamefont {Vernizzi}},\ }\href
  {\doibase 10.1088/1475-7516/2016/02/056} {\bibfield  {journal} {\bibinfo
  {journal} {JCAP}\ }\textbf {\bibinfo {volume} {1602}},\ \bibinfo {pages}
  {056} (\bibinfo {year} {2016})},\ \Eprint {http://arxiv.org/abs/1509.02191}
  {arXiv:1509.02191 [astro-ph.CO]} \BibitemShut {NoStop}%
\bibitem [{\citenamefont {Bellini}\ \emph {et~al.}(2016)\citenamefont
  {Bellini}, \citenamefont {Cuesta}, \citenamefont {Jim{\'e}nez},\ and\
  \citenamefont {Verde}}]{Bellini:2015xja}%
  \BibitemOpen
  \bibfield  {author} {\bibinfo {author} {\bibfnamefont {E.}~\bibnamefont
  {Bellini}}, \bibinfo {author} {\bibfnamefont {A.~J.}\ \bibnamefont {Cuesta}},
  \bibinfo {author} {\bibfnamefont {R.}~\bibnamefont {Jim{\'e}nez}}, \ and\
  \bibinfo {author} {\bibfnamefont {L.}~\bibnamefont {Verde}},\ }\href
  {\doibase 10.1088/1475-7516/2016/06/E01, 10.1088/1475-7516/2016/02/053}
  {\bibfield  {journal} {\bibinfo  {journal} {JCAP}\ }\textbf {\bibinfo
  {volume} {1602}},\ \bibinfo {pages} {053} (\bibinfo {year} {2016})},\
  \bibinfo {note} {[Erratum: JCAP1606,no.06,E01(2016)]},\ \Eprint
  {http://arxiv.org/abs/1509.07816} {arXiv:1509.07816 [astro-ph.CO]}
  \BibitemShut {NoStop}%
\bibitem [{\citenamefont {Alonso}\ \emph {et~al.}(2017)\citenamefont {Alonso},
  \citenamefont {Bellini}, \citenamefont {Ferreira},\ and\ \citenamefont
  {Zumalac{\'a}rregui}}]{Alonso:2016suf}%
  \BibitemOpen
  \bibfield  {author} {\bibinfo {author} {\bibfnamefont {D.}~\bibnamefont
  {Alonso}}, \bibinfo {author} {\bibfnamefont {E.}~\bibnamefont {Bellini}},
  \bibinfo {author} {\bibfnamefont {P.~G.}\ \bibnamefont {Ferreira}}, \ and\
  \bibinfo {author} {\bibfnamefont {M.}~\bibnamefont {Zumalac{\'a}rregui}},\
  }\href {\doibase 10.1103/PhysRevD.95.063502} {\bibfield  {journal} {\bibinfo
  {journal} {Phys. Rev.}\ }\textbf {\bibinfo {volume} {D95}},\ \bibinfo {pages}
  {063502} (\bibinfo {year} {2017})},\ \Eprint
  {http://arxiv.org/abs/1610.09290} {arXiv:1610.09290 [astro-ph.CO]}
  \BibitemShut {NoStop}%
\bibitem [{\citenamefont {Noller}\ and\ \citenamefont
  {Nicola}(2019)}]{Noller:2018wyv}%
  \BibitemOpen
  \bibfield  {author} {\bibinfo {author} {\bibfnamefont {J.}~\bibnamefont
  {Noller}}\ and\ \bibinfo {author} {\bibfnamefont {A.}~\bibnamefont
  {Nicola}},\ }\href {\doibase 10.1103/PhysRevD.99.103502} {\bibfield
  {journal} {\bibinfo  {journal} {Phys. Rev.}\ }\textbf {\bibinfo {volume}
  {D99}},\ \bibinfo {pages} {103502} (\bibinfo {year} {2019})},\ \Eprint
  {http://arxiv.org/abs/1811.12928} {arXiv:1811.12928 [astro-ph.CO]}
  \BibitemShut {NoStop}%
\bibitem [{\citenamefont {Noller}\ and\ \citenamefont
  {Nicola}(2018)}]{Noller:2018eht}%
  \BibitemOpen
  \bibfield  {author} {\bibinfo {author} {\bibfnamefont {J.}~\bibnamefont
  {Noller}}\ and\ \bibinfo {author} {\bibfnamefont {A.}~\bibnamefont
  {Nicola}},\ }\href@noop {} {\  (\bibinfo {year} {2018})},\ \Eprint
  {http://arxiv.org/abs/1811.03082} {arXiv:1811.03082 [astro-ph.CO]}
  \BibitemShut {NoStop}%
\bibitem [{\citenamefont {Bonvin}\ and\ \citenamefont
  {Fleury}(2018)}]{Bonvin:2018ckp}%
  \BibitemOpen
  \bibfield  {author} {\bibinfo {author} {\bibfnamefont {C.}~\bibnamefont
  {Bonvin}}\ and\ \bibinfo {author} {\bibfnamefont {P.}~\bibnamefont
  {Fleury}},\ }\href {\doibase 10.1088/1475-7516/2018/05/061} {\bibfield
  {journal} {\bibinfo  {journal} {JCAP}\ }\textbf {\bibinfo {volume} {1805}},\
  \bibinfo {pages} {061} (\bibinfo {year} {2018})},\ \Eprint
  {http://arxiv.org/abs/1803.02771} {arXiv:1803.02771 [astro-ph.CO]}
  \BibitemShut {NoStop}%
\bibitem [{\citenamefont {Desmond}\ \emph {et~al.}(2019)\citenamefont
  {Desmond}, \citenamefont {Ferreira}, \citenamefont {Lavaux},\ and\
  \citenamefont {Jasche}}]{Desmond:2018euk}%
  \BibitemOpen
  \bibfield  {author} {\bibinfo {author} {\bibfnamefont {H.}~\bibnamefont
  {Desmond}}, \bibinfo {author} {\bibfnamefont {P.~G.}\ \bibnamefont
  {Ferreira}}, \bibinfo {author} {\bibfnamefont {G.}~\bibnamefont {Lavaux}}, \
  and\ \bibinfo {author} {\bibfnamefont {J.}~\bibnamefont {Jasche}},\ }\href
  {\doibase 10.1093/mnrasl/sly221} {\bibfield  {journal} {\bibinfo  {journal}
  {Mon. Not. Roy. Astron. Soc.}\ }\textbf {\bibinfo {volume} {483}},\ \bibinfo
  {pages} {L64} (\bibinfo {year} {2019})},\ \Eprint
  {http://arxiv.org/abs/1802.07206} {arXiv:1802.07206 [astro-ph.CO]}
  \BibitemShut {NoStop}%
\bibitem [{\citenamefont {Abbott}\ \emph {et~al.}(2017)\citenamefont {Abbott}
  \emph {et~al.}}]{TheLIGOScientific:2017qsa}%
  \BibitemOpen
  \bibfield  {author} {\bibinfo {author} {\bibfnamefont {B.~P.}\ \bibnamefont
  {Abbott}} \emph {et~al.} (\bibinfo {collaboration} {Virgo, LIGO
  Scientific}),\ }\href {\doibase 10.1103/PhysRevLett.119.161101} {\bibfield
  {journal} {\bibinfo  {journal} {Phys. Rev. Lett.}\ }\textbf {\bibinfo
  {volume} {119}},\ \bibinfo {pages} {161101} (\bibinfo {year} {2017})},\
  \Eprint {http://arxiv.org/abs/1710.05832} {arXiv:1710.05832 [gr-qc]}
  \BibitemShut {NoStop}%
\bibitem [{\citenamefont {Goldstein}\ \emph {et~al.}(2017)\citenamefont
  {Goldstein} \emph {et~al.}}]{Goldstein:2017mmi}%
  \BibitemOpen
  \bibfield  {author} {\bibinfo {author} {\bibfnamefont {A.}~\bibnamefont
  {Goldstein}} \emph {et~al.},\ }\href {\doibase 10.3847/2041-8213/aa8f41}
  {\bibfield  {journal} {\bibinfo  {journal} {Astrophys. J.}\ }\textbf
  {\bibinfo {volume} {848}},\ \bibinfo {pages} {L14} (\bibinfo {year}
  {2017})},\ \Eprint {http://arxiv.org/abs/1710.05446} {arXiv:1710.05446
  [astro-ph.HE]} \BibitemShut {NoStop}%
\bibitem [{\citenamefont {Savchenko}\ \emph {et~al.}(2017)\citenamefont
  {Savchenko} \emph {et~al.}}]{Savchenko:2017ffs}%
  \BibitemOpen
  \bibfield  {author} {\bibinfo {author} {\bibfnamefont {V.}~\bibnamefont
  {Savchenko}} \emph {et~al.},\ }\href {\doibase 10.3847/2041-8213/aa8f94}
  {\bibfield  {journal} {\bibinfo  {journal} {Astrophys. J.}\ }\textbf
  {\bibinfo {volume} {848}},\ \bibinfo {pages} {L15} (\bibinfo {year}
  {2017})},\ \Eprint {http://arxiv.org/abs/1710.05449} {arXiv:1710.05449
  [astro-ph.HE]} \BibitemShut {NoStop}%
\bibitem [{\citenamefont {Kimura}\ and\ \citenamefont
  {Yamamoto}(2012)}]{Kimura:2011qn}%
  \BibitemOpen
  \bibfield  {author} {\bibinfo {author} {\bibfnamefont {R.}~\bibnamefont
  {Kimura}}\ and\ \bibinfo {author} {\bibfnamefont {K.}~\bibnamefont
  {Yamamoto}},\ }\href {\doibase 10.1088/1475-7516/2012/07/050} {\bibfield
  {journal} {\bibinfo  {journal} {JCAP}\ }\textbf {\bibinfo {volume} {1207}},\
  \bibinfo {pages} {050} (\bibinfo {year} {2012})},\ \Eprint
  {http://arxiv.org/abs/1112.4284} {arXiv:1112.4284 [astro-ph.CO]} \BibitemShut
  {NoStop}%
\bibitem [{\citenamefont {McManus}\ \emph {et~al.}(2016)\citenamefont
  {McManus}, \citenamefont {Lombriser},\ and\ \citenamefont
  {Pe{\~n}arrubia}}]{McManus:2016kxu}%
  \BibitemOpen
  \bibfield  {author} {\bibinfo {author} {\bibfnamefont {R.}~\bibnamefont
  {McManus}}, \bibinfo {author} {\bibfnamefont {L.}~\bibnamefont {Lombriser}},
  \ and\ \bibinfo {author} {\bibfnamefont {J.}~\bibnamefont {Pe{\~n}arrubia}},\
  }\href {\doibase 10.1088/1475-7516/2016/11/006} {\bibfield  {journal}
  {\bibinfo  {journal} {JCAP}\ }\textbf {\bibinfo {volume} {1611}},\ \bibinfo
  {pages} {006} (\bibinfo {year} {2016})},\ \Eprint
  {http://arxiv.org/abs/1606.03282} {arXiv:1606.03282 [gr-qc]} \BibitemShut
  {NoStop}%
\bibitem [{\citenamefont {Lombriser}\ and\ \citenamefont
  {Taylor}(2016)}]{Lombriser:2015sxa}%
  \BibitemOpen
  \bibfield  {author} {\bibinfo {author} {\bibfnamefont {L.}~\bibnamefont
  {Lombriser}}\ and\ \bibinfo {author} {\bibfnamefont {A.}~\bibnamefont
  {Taylor}},\ }\href {\doibase 10.1088/1475-7516/2016/03/031} {\bibfield
  {journal} {\bibinfo  {journal} {JCAP}\ }\textbf {\bibinfo {volume} {1603}},\
  \bibinfo {pages} {031} (\bibinfo {year} {2016})},\ \Eprint
  {http://arxiv.org/abs/1509.08458} {arXiv:1509.08458 [astro-ph.CO]}
  \BibitemShut {NoStop}%
\bibitem [{\citenamefont {Lombriser}\ and\ \citenamefont
  {Lima}(2017)}]{Lombriser:2016yzn}%
  \BibitemOpen
  \bibfield  {author} {\bibinfo {author} {\bibfnamefont {L.}~\bibnamefont
  {Lombriser}}\ and\ \bibinfo {author} {\bibfnamefont {N.~A.}\ \bibnamefont
  {Lima}},\ }\href {\doibase 10.1016/j.physletb.2016.12.048} {\bibfield
  {journal} {\bibinfo  {journal} {Phys. Lett.}\ }\textbf {\bibinfo {volume}
  {B765}},\ \bibinfo {pages} {382} (\bibinfo {year} {2017})},\ \Eprint
  {http://arxiv.org/abs/1602.07670} {arXiv:1602.07670 [astro-ph.CO]}
  \BibitemShut {NoStop}%
\bibitem [{\citenamefont {Ezquiaga}\ and\ \citenamefont
  {Zumalac{\'a}rregui}(2017)}]{Ezquiaga:2017ekz}%
  \BibitemOpen
  \bibfield  {author} {\bibinfo {author} {\bibfnamefont {J.~M.}\ \bibnamefont
  {Ezquiaga}}\ and\ \bibinfo {author} {\bibfnamefont {M.}~\bibnamefont
  {Zumalac{\'a}rregui}},\ }\href {\doibase 10.1103/PhysRevLett.119.251304}
  {\bibfield  {journal} {\bibinfo  {journal} {Phys. Rev. Lett.}\ }\textbf
  {\bibinfo {volume} {119}},\ \bibinfo {pages} {251304} (\bibinfo {year}
  {2017})},\ \Eprint {http://arxiv.org/abs/1710.05901} {arXiv:1710.05901
  [astro-ph.CO]} \BibitemShut {NoStop}%
\bibitem [{\citenamefont {Creminelli}\ and\ \citenamefont
  {Vernizzi}(2017)}]{Creminelli:2017sry}%
  \BibitemOpen
  \bibfield  {author} {\bibinfo {author} {\bibfnamefont {P.}~\bibnamefont
  {Creminelli}}\ and\ \bibinfo {author} {\bibfnamefont {F.}~\bibnamefont
  {Vernizzi}},\ }\href {\doibase 10.1103/PhysRevLett.119.251302} {\bibfield
  {journal} {\bibinfo  {journal} {Phys. Rev. Lett.}\ }\textbf {\bibinfo
  {volume} {119}},\ \bibinfo {pages} {251302} (\bibinfo {year} {2017})},\
  \Eprint {http://arxiv.org/abs/1710.05877} {arXiv:1710.05877 [astro-ph.CO]}
  \BibitemShut {NoStop}%
\bibitem [{\citenamefont {Sakstein}\ and\ \citenamefont
  {Jain}(2017)}]{Sakstein:2017xjx}%
  \BibitemOpen
  \bibfield  {author} {\bibinfo {author} {\bibfnamefont {J.}~\bibnamefont
  {Sakstein}}\ and\ \bibinfo {author} {\bibfnamefont {B.}~\bibnamefont
  {Jain}},\ }\href {\doibase 10.1103/PhysRevLett.119.251303} {\bibfield
  {journal} {\bibinfo  {journal} {Phys. Rev. Lett.}\ }\textbf {\bibinfo
  {volume} {119}},\ \bibinfo {pages} {251303} (\bibinfo {year} {2017})},\
  \Eprint {http://arxiv.org/abs/1710.05893} {arXiv:1710.05893 [astro-ph.CO]}
  \BibitemShut {NoStop}%
\bibitem [{\citenamefont {Langlois}\ \emph {et~al.}(2018)\citenamefont
  {Langlois}, \citenamefont {Saito}, \citenamefont {Yamauchi},\ and\
  \citenamefont {Noui}}]{Langlois:2017}%
  \BibitemOpen
  \bibfield  {author} {\bibinfo {author} {\bibfnamefont {D.}~\bibnamefont
  {Langlois}}, \bibinfo {author} {\bibfnamefont {R.}~\bibnamefont {Saito}},
  \bibinfo {author} {\bibfnamefont {D.}~\bibnamefont {Yamauchi}}, \ and\
  \bibinfo {author} {\bibfnamefont {K.}~\bibnamefont {Noui}},\ }\href {\doibase
  10.1103/PhysRevD.97.061501} {\bibfield  {journal} {\bibinfo  {journal} {Phys.
  Rev. D}\ }\textbf {\bibinfo {volume} {97}},\ \bibinfo {pages} {061501}
  (\bibinfo {year} {2018})}\BibitemShut {NoStop}%
\bibitem [{\citenamefont {Baker}\ \emph {et~al.}(2017)\citenamefont {Baker},
  \citenamefont {Bellini}, \citenamefont {Ferreira}, \citenamefont {Lagos},
  \citenamefont {Noller},\ and\ \citenamefont {Sawicki}}]{Baker:2017hug}%
  \BibitemOpen
  \bibfield  {author} {\bibinfo {author} {\bibfnamefont {T.}~\bibnamefont
  {Baker}}, \bibinfo {author} {\bibfnamefont {E.}~\bibnamefont {Bellini}},
  \bibinfo {author} {\bibfnamefont {P.~G.}\ \bibnamefont {Ferreira}}, \bibinfo
  {author} {\bibfnamefont {M.}~\bibnamefont {Lagos}}, \bibinfo {author}
  {\bibfnamefont {J.}~\bibnamefont {Noller}}, \ and\ \bibinfo {author}
  {\bibfnamefont {I.}~\bibnamefont {Sawicki}},\ }\href {\doibase
  10.1103/PhysRevLett.119.251301} {\bibfield  {journal} {\bibinfo  {journal}
  {Phys. Rev. Lett.}\ }\textbf {\bibinfo {volume} {119}},\ \bibinfo {pages}
  {251301} (\bibinfo {year} {2017})},\ \Eprint
  {http://arxiv.org/abs/1710.06394} {arXiv:1710.06394 [astro-ph.CO]}
  \BibitemShut {NoStop}%
\bibitem [{\citenamefont {Holz}\ and\ \citenamefont
  {Hughes}(2005)}]{Holz:2005df}%
  \BibitemOpen
  \bibfield  {author} {\bibinfo {author} {\bibfnamefont {D.~E.}\ \bibnamefont
  {Holz}}\ and\ \bibinfo {author} {\bibfnamefont {S.~A.}\ \bibnamefont
  {Hughes}},\ }\href {\doibase 10.1086/431341} {\bibfield  {journal} {\bibinfo
  {journal} {Astrophys. J.}\ }\textbf {\bibinfo {volume} {629}},\ \bibinfo
  {pages} {15} (\bibinfo {year} {2005})},\ \Eprint
  {http://arxiv.org/abs/astro-ph/0504616} {arXiv:astro-ph/0504616 [astro-ph]}
  \BibitemShut {NoStop}%
\bibitem [{\citenamefont {{Schutz}}(1986)}]{1986Natur.323..310S}%
  \BibitemOpen
  \bibfield  {author} {\bibinfo {author} {\bibfnamefont {B.~F.}\ \bibnamefont
  {{Schutz}}},\ }\href {\doibase 10.1038/323310a0} {\bibfield  {journal}
  {\bibinfo  {journal} {\nat}\ }\textbf {\bibinfo {volume} {323}},\ \bibinfo
  {pages} {310} (\bibinfo {year} {1986})}\BibitemShut {NoStop}%
\bibitem [{\citenamefont {Chen}\ \emph {et~al.}(2018)\citenamefont {Chen},
  \citenamefont {Fishbach},\ and\ \citenamefont {Holz}}]{Chen:2017rfc}%
  \BibitemOpen
  \bibfield  {author} {\bibinfo {author} {\bibfnamefont {H.-Y.}\ \bibnamefont
  {Chen}}, \bibinfo {author} {\bibfnamefont {M.}~\bibnamefont {Fishbach}}, \
  and\ \bibinfo {author} {\bibfnamefont {D.~E.}\ \bibnamefont {Holz}},\ }\href
  {\doibase 10.1038/s41586-018-0606-0} {\bibfield  {journal} {\bibinfo
  {journal} {Nature}\ }\textbf {\bibinfo {volume} {562}},\ \bibinfo {pages}
  {545} (\bibinfo {year} {2018})},\ \Eprint {http://arxiv.org/abs/1712.06531}
  {arXiv:1712.06531 [astro-ph.CO]} \BibitemShut {NoStop}%
\bibitem [{\citenamefont {Soares-Santos}\ \emph {et~al.}(2019)\citenamefont
  {Soares-Santos} \emph {et~al.}}]{Soares-Santos:2019irc}%
  \BibitemOpen
  \bibfield  {author} {\bibinfo {author} {\bibfnamefont {M.}~\bibnamefont
  {Soares-Santos}} \emph {et~al.} (\bibinfo {collaboration} {DES, LIGO
  Scientific, Virgo}),\ }\href {\doibase 10.3847/2041-8213/ab14f1} {\bibfield
  {journal} {\bibinfo  {journal} {Astrophys. J. Lett.}\ }\textbf {\bibinfo
  {volume} {876}},\ \bibinfo {pages} {L7} (\bibinfo {year} {2019})},\ \Eprint
  {http://arxiv.org/abs/1901.01540} {arXiv:1901.01540 [astro-ph.CO]}
  \BibitemShut {NoStop}%
\bibitem [{\citenamefont {Saltas}\ \emph {et~al.}(2014)\citenamefont {Saltas},
  \citenamefont {Sawicki}, \citenamefont {Amendola},\ and\ \citenamefont
  {Kunz}}]{Saltas:2014dha}%
  \BibitemOpen
  \bibfield  {author} {\bibinfo {author} {\bibfnamefont {I.~D.}\ \bibnamefont
  {Saltas}}, \bibinfo {author} {\bibfnamefont {I.}~\bibnamefont {Sawicki}},
  \bibinfo {author} {\bibfnamefont {L.}~\bibnamefont {Amendola}}, \ and\
  \bibinfo {author} {\bibfnamefont {M.}~\bibnamefont {Kunz}},\ }\href {\doibase
  10.1103/PhysRevLett.113.191101} {\bibfield  {journal} {\bibinfo  {journal}
  {Phys. Rev. Lett.}\ }\textbf {\bibinfo {volume} {113}},\ \bibinfo {pages}
  {191101} (\bibinfo {year} {2014})},\ \Eprint {http://arxiv.org/abs/1406.7139}
  {arXiv:1406.7139 [astro-ph.CO]} \BibitemShut {NoStop}%
\bibitem [{\citenamefont {Amendola}\ \emph {et~al.}(2018)\citenamefont
  {Amendola}, \citenamefont {Sawicki}, \citenamefont {Kunz},\ and\
  \citenamefont {Saltas}}]{Amendola:2017ovw}%
  \BibitemOpen
  \bibfield  {author} {\bibinfo {author} {\bibfnamefont {L.}~\bibnamefont
  {Amendola}}, \bibinfo {author} {\bibfnamefont {I.}~\bibnamefont {Sawicki}},
  \bibinfo {author} {\bibfnamefont {M.}~\bibnamefont {Kunz}}, \ and\ \bibinfo
  {author} {\bibfnamefont {I.~D.}\ \bibnamefont {Saltas}},\ }\href {\doibase
  10.1088/1475-7516/2018/08/030} {\bibfield  {journal} {\bibinfo  {journal}
  {JCAP}\ }\textbf {\bibinfo {volume} {1808}},\ \bibinfo {pages} {030}
  (\bibinfo {year} {2018})},\ \Eprint {http://arxiv.org/abs/1712.08623}
  {arXiv:1712.08623 [astro-ph.CO]} \BibitemShut {NoStop}%
\bibitem [{\citenamefont {Belgacem}\ \emph
  {et~al.}(2018{\natexlab{a}})\citenamefont {Belgacem}, \citenamefont {Dirian},
  \citenamefont {Foffa},\ and\ \citenamefont {Maggiore}}]{Belgacem:2017ihm}%
  \BibitemOpen
  \bibfield  {author} {\bibinfo {author} {\bibfnamefont {E.}~\bibnamefont
  {Belgacem}}, \bibinfo {author} {\bibfnamefont {Y.}~\bibnamefont {Dirian}},
  \bibinfo {author} {\bibfnamefont {S.}~\bibnamefont {Foffa}}, \ and\ \bibinfo
  {author} {\bibfnamefont {M.}~\bibnamefont {Maggiore}},\ }\href {\doibase
  10.1103/PhysRevD.97.104066} {\bibfield  {journal} {\bibinfo  {journal} {Phys.
  Rev.}\ }\textbf {\bibinfo {volume} {D97}},\ \bibinfo {pages} {104066}
  (\bibinfo {year} {2018}{\natexlab{a}})},\ \Eprint
  {http://arxiv.org/abs/1712.08108} {arXiv:1712.08108 [astro-ph.CO]}
  \BibitemShut {NoStop}%
\bibitem [{\citenamefont {Belgacem}\ \emph
  {et~al.}(2018{\natexlab{b}})\citenamefont {Belgacem}, \citenamefont {Dirian},
  \citenamefont {Foffa},\ and\ \citenamefont {Maggiore}}]{Belgacem:2018lbp}%
  \BibitemOpen
  \bibfield  {author} {\bibinfo {author} {\bibfnamefont {E.}~\bibnamefont
  {Belgacem}}, \bibinfo {author} {\bibfnamefont {Y.}~\bibnamefont {Dirian}},
  \bibinfo {author} {\bibfnamefont {S.}~\bibnamefont {Foffa}}, \ and\ \bibinfo
  {author} {\bibfnamefont {M.}~\bibnamefont {Maggiore}},\ }\href {\doibase
  10.1103/PhysRevD.98.023510} {\bibfield  {journal} {\bibinfo  {journal} {Phys.
  Rev.}\ }\textbf {\bibinfo {volume} {D98}},\ \bibinfo {pages} {023510}
  (\bibinfo {year} {2018}{\natexlab{b}})},\ \Eprint
  {http://arxiv.org/abs/1805.08731} {arXiv:1805.08731 [gr-qc]} \BibitemShut
  {NoStop}%
\bibitem [{\citenamefont {Lagos}\ \emph {et~al.}(2019)\citenamefont {Lagos},
  \citenamefont {Fishbach}, \citenamefont {Landry},\ and\ \citenamefont
  {Holz}}]{Lagos:2019kds}%
  \BibitemOpen
  \bibfield  {author} {\bibinfo {author} {\bibfnamefont {M.}~\bibnamefont
  {Lagos}}, \bibinfo {author} {\bibfnamefont {M.}~\bibnamefont {Fishbach}},
  \bibinfo {author} {\bibfnamefont {P.}~\bibnamefont {Landry}}, \ and\ \bibinfo
  {author} {\bibfnamefont {D.~E.}\ \bibnamefont {Holz}},\ }\href {\doibase
  10.1103/PhysRevD.99.083504} {\bibfield  {journal} {\bibinfo  {journal} {Phys.
  Rev.}\ }\textbf {\bibinfo {volume} {D99}},\ \bibinfo {pages} {083504}
  (\bibinfo {year} {2019})},\ \Eprint {http://arxiv.org/abs/1901.03321}
  {arXiv:1901.03321 [astro-ph.CO]} \BibitemShut {NoStop}%
\bibitem [{\citenamefont {D'Agostino}\ and\ \citenamefont
  {Nunes}(2019)}]{DAgostino:2019hvh}%
  \BibitemOpen
  \bibfield  {author} {\bibinfo {author} {\bibfnamefont {R.}~\bibnamefont
  {D'Agostino}}\ and\ \bibinfo {author} {\bibfnamefont {R.~C.}\ \bibnamefont
  {Nunes}},\ }\href {\doibase 10.1103/PhysRevD.100.044041} {\bibfield
  {journal} {\bibinfo  {journal} {Phys. Rev.}\ }\textbf {\bibinfo {volume}
  {D100}},\ \bibinfo {pages} {044041} (\bibinfo {year} {2019})},\ \Eprint
  {http://arxiv.org/abs/1907.05516} {arXiv:1907.05516 [gr-qc]} \BibitemShut
  {NoStop}%
\bibitem [{Lis()}]{Lisa:2020}%
  \BibitemOpen
  \href@noop {} {\emph {\bibinfo {title} {LISA}}},\ \bibinfo {note}
  {\url{https://www.elisascience.org}}\BibitemShut {NoStop}%
\bibitem [{\citenamefont {Belgacem}\ \emph {et~al.}(2019)\citenamefont
  {Belgacem} \emph {et~al.}}]{Belgacem:2019pkk}%
  \BibitemOpen
  \bibfield  {author} {\bibinfo {author} {\bibfnamefont {E.}~\bibnamefont
  {Belgacem}} \emph {et~al.} (\bibinfo {collaboration} {LISA Cosmology Working
  Group}),\ }\href {\doibase 10.1088/1475-7516/2019/07/024} {\bibfield
  {journal} {\bibinfo  {journal} {JCAP}\ }\textbf {\bibinfo {volume} {1907}},\
  \bibinfo {pages} {024} (\bibinfo {year} {2019})},\ \Eprint
  {http://arxiv.org/abs/1906.01593} {arXiv:1906.01593 [astro-ph.CO]}
  \BibitemShut {NoStop}%
\bibitem [{\citenamefont {Dalang}\ and\ \citenamefont
  {Lombriser}(2019)}]{Dalang:2019fma}%
  \BibitemOpen
  \bibfield  {author} {\bibinfo {author} {\bibfnamefont {C.}~\bibnamefont
  {Dalang}}\ and\ \bibinfo {author} {\bibfnamefont {L.}~\bibnamefont
  {Lombriser}},\ }\href {\doibase 10.1088/1475-7516/2019/10/013} {\bibfield
  {journal} {\bibinfo  {journal} {JCAP}\ }\textbf {\bibinfo {volume} {1910}},\
  \bibinfo {pages} {013} (\bibinfo {year} {2019})},\ \Eprint
  {http://arxiv.org/abs/1906.12333} {arXiv:1906.12333 [astro-ph.CO]}
  \BibitemShut {NoStop}%
\bibitem [{\citenamefont {Misner}\ \emph {et~al.}(1973)\citenamefont {Misner},
  \citenamefont {Thorne},\ and\ \citenamefont {Wheeler}}]{Misner:1974qy}%
  \BibitemOpen
  \bibfield  {author} {\bibinfo {author} {\bibfnamefont {C.~W.}\ \bibnamefont
  {Misner}}, \bibinfo {author} {\bibfnamefont {K.~S.}\ \bibnamefont {Thorne}},
  \ and\ \bibinfo {author} {\bibfnamefont {J.~A.}\ \bibnamefont {Wheeler}},\
  }\href@noop {} {\emph {\bibinfo {title} {{Gravitation}}}}\ (\bibinfo
  {publisher} {W. H. Freeman},\ \bibinfo {address} {San Francisco},\ \bibinfo
  {year} {1973})\BibitemShut {NoStop}%
\bibitem [{\citenamefont {Tsujikawa}(2013)}]{Tsujikawa:2013fta}%
  \BibitemOpen
  \bibfield  {author} {\bibinfo {author} {\bibfnamefont {S.}~\bibnamefont
  {Tsujikawa}},\ }\href {\doibase 10.1088/0264-9381/30/21/214003} {\bibfield
  {journal} {\bibinfo  {journal} {Class. Quant. Grav.}\ }\textbf {\bibinfo
  {volume} {30}},\ \bibinfo {pages} {214003} (\bibinfo {year} {2013})},\
  \Eprint {http://arxiv.org/abs/1304.1961} {arXiv:1304.1961 [gr-qc]}
  \BibitemShut {NoStop}%
\bibitem [{\citenamefont {Brans}\ and\ \citenamefont
  {Dicke}(1961)}]{Brans:1961sx}%
  \BibitemOpen
  \bibfield  {author} {\bibinfo {author} {\bibfnamefont {C.}~\bibnamefont
  {Brans}}\ and\ \bibinfo {author} {\bibfnamefont {R.~H.}\ \bibnamefont
  {Dicke}},\ }\href {\doibase 10.1103/PhysRev.124.925} {\bibfield  {journal}
  {\bibinfo  {journal} {Phys. Rev.}\ }\textbf {\bibinfo {volume} {124}},\
  \bibinfo {pages} {925} (\bibinfo {year} {1961})},\ \bibinfo {note}
  {[,142(1961)]}\BibitemShut {NoStop}%
\bibitem [{\citenamefont {De~Felice}\ and\ \citenamefont
  {Tsujikawa}(2010)}]{DeFelice:2010aj}%
  \BibitemOpen
  \bibfield  {author} {\bibinfo {author} {\bibfnamefont {A.}~\bibnamefont
  {De~Felice}}\ and\ \bibinfo {author} {\bibfnamefont {S.}~\bibnamefont
  {Tsujikawa}},\ }\href {\doibase 10.12942/lrr-2010-3} {\bibfield  {journal}
  {\bibinfo  {journal} {Living Rev. Rel.}\ }\textbf {\bibinfo {volume} {13}},\
  \bibinfo {pages} {3} (\bibinfo {year} {2010})},\ \Eprint
  {http://arxiv.org/abs/1002.4928} {arXiv:1002.4928 [gr-qc]} \BibitemShut
  {NoStop}%
\bibitem [{\citenamefont {Deffayet}\ \emph {et~al.}(2009)\citenamefont
  {Deffayet}, \citenamefont {Esposito-Far{\`e}se},\ and\ \citenamefont
  {Vikman}}]{Deffayet:2009wt}%
  \BibitemOpen
  \bibfield  {author} {\bibinfo {author} {\bibfnamefont {C.}~\bibnamefont
  {Deffayet}}, \bibinfo {author} {\bibfnamefont {G.}~\bibnamefont
  {Esposito-Far{\`e}se}}, \ and\ \bibinfo {author} {\bibfnamefont
  {A.}~\bibnamefont {Vikman}},\ }\href {\doibase 10.1103/PhysRevD.79.084003}
  {\bibfield  {journal} {\bibinfo  {journal} {Phys. Rev.}\ }\textbf {\bibinfo
  {volume} {D79}},\ \bibinfo {pages} {084003} (\bibinfo {year} {2009})},\
  \Eprint {http://arxiv.org/abs/0901.1314} {arXiv:0901.1314 [hep-th]}
  \BibitemShut {NoStop}%
\bibitem [{\citenamefont {Perenon}\ \emph {et~al.}(2015)\citenamefont
  {Perenon}, \citenamefont {Piazza}, \citenamefont {Marinoni},\ and\
  \citenamefont {Hui}}]{Perenon:2015sla}%
  \BibitemOpen
  \bibfield  {author} {\bibinfo {author} {\bibfnamefont {L.}~\bibnamefont
  {Perenon}}, \bibinfo {author} {\bibfnamefont {F.}~\bibnamefont {Piazza}},
  \bibinfo {author} {\bibfnamefont {C.}~\bibnamefont {Marinoni}}, \ and\
  \bibinfo {author} {\bibfnamefont {L.}~\bibnamefont {Hui}},\ }\href {\doibase
  10.1088/1475-7516/2015/11/029} {\bibfield  {journal} {\bibinfo  {journal}
  {JCAP}\ }\textbf {\bibinfo {volume} {1511}},\ \bibinfo {pages} {029}
  (\bibinfo {year} {2015})},\ \Eprint {http://arxiv.org/abs/1506.03047}
  {arXiv:1506.03047 [astro-ph.CO]} \BibitemShut {NoStop}%
\bibitem [{\citenamefont {Bettoni}\ and\ \citenamefont
  {Liberati}(2013)}]{Bettoni:2013diz}%
  \BibitemOpen
  \bibfield  {author} {\bibinfo {author} {\bibfnamefont {D.}~\bibnamefont
  {Bettoni}}\ and\ \bibinfo {author} {\bibfnamefont {S.}~\bibnamefont
  {Liberati}},\ }\href {\doibase 10.1103/PhysRevD.88.084020} {\bibfield
  {journal} {\bibinfo  {journal} {Phys. Rev.}\ }\textbf {\bibinfo {volume}
  {D88}},\ \bibinfo {pages} {084020} (\bibinfo {year} {2013})},\ \Eprint
  {http://arxiv.org/abs/1306.6724} {arXiv:1306.6724 [gr-qc]} \BibitemShut
  {NoStop}%
\bibitem [{\citenamefont {Francfort}\ \emph {et~al.}(2019)\citenamefont
  {Francfort}, \citenamefont {Ghosh},\ and\ \citenamefont
  {Durrer}}]{Francfort:2019ynz}%
  \BibitemOpen
  \bibfield  {author} {\bibinfo {author} {\bibfnamefont {J.}~\bibnamefont
  {Francfort}}, \bibinfo {author} {\bibfnamefont {B.}~\bibnamefont {Ghosh}}, \
  and\ \bibinfo {author} {\bibfnamefont {R.}~\bibnamefont {Durrer}},\ }\href
  {\doibase 10.1088/1475-7516/2019/09/071} {\bibfield  {journal} {\bibinfo
  {journal} {JCAP}\ }\textbf {\bibinfo {volume} {1909}},\ \bibinfo {pages}
  {071} (\bibinfo {year} {2019})},\ \Eprint {http://arxiv.org/abs/1907.03606}
  {arXiv:1907.03606 [gr-qc]} \BibitemShut {NoStop}%
\bibitem [{Lig()}]{Ligo:2020}%
  \BibitemOpen
  \href@noop {} {\emph {\bibinfo {title} {LIGO}}},\ \bibinfo {note}
  {\url{https://www.ligo.caltech.edu}}\BibitemShut {NoStop}%
\bibitem [{Vir()}]{Virgo:2020}%
  \BibitemOpen
  \href@noop {} {\emph {\bibinfo {title} {VIRGO}}},\ \bibinfo {note}
  {\url{http://www.virgo-gw.eu/}}\BibitemShut {NoStop}%
\bibitem [{\citenamefont {Copeland}\ \emph {et~al.}(2019)\citenamefont
  {Copeland}, \citenamefont {Kopp}, \citenamefont {Padilla}, \citenamefont
  {Saffin},\ and\ \citenamefont {Skordis}}]{Copeland:2018yuh}%
  \BibitemOpen
  \bibfield  {author} {\bibinfo {author} {\bibfnamefont {E.~J.}\ \bibnamefont
  {Copeland}}, \bibinfo {author} {\bibfnamefont {M.}~\bibnamefont {Kopp}},
  \bibinfo {author} {\bibfnamefont {A.}~\bibnamefont {Padilla}}, \bibinfo
  {author} {\bibfnamefont {P.~M.}\ \bibnamefont {Saffin}}, \ and\ \bibinfo
  {author} {\bibfnamefont {C.}~\bibnamefont {Skordis}},\ }\href {\doibase
  10.1103/PhysRevLett.122.061301} {\bibfield  {journal} {\bibinfo  {journal}
  {Phys. Rev. Lett.}\ }\textbf {\bibinfo {volume} {122}},\ \bibinfo {pages}
  {061301} (\bibinfo {year} {2019})},\ \Eprint
  {http://arxiv.org/abs/1810.08239} {arXiv:1810.08239 [gr-qc]} \BibitemShut
  {NoStop}%
\bibitem [{\citenamefont {de~Rham}\ and\ \citenamefont
  {Melville}(2018)}]{deRham:2018red}%
  \BibitemOpen
  \bibfield  {author} {\bibinfo {author} {\bibfnamefont {C.}~\bibnamefont
  {de~Rham}}\ and\ \bibinfo {author} {\bibfnamefont {S.}~\bibnamefont
  {Melville}},\ }\href {\doibase 10.1103/PhysRevLett.121.221101} {\bibfield
  {journal} {\bibinfo  {journal} {Phys. Rev. Lett.}\ }\textbf {\bibinfo
  {volume} {121}},\ \bibinfo {pages} {221101} (\bibinfo {year} {2018})},\
  \Eprint {http://arxiv.org/abs/1806.09417} {arXiv:1806.09417 [hep-th]}
  \BibitemShut {NoStop}%
\bibitem [{\citenamefont {Bonilla}\ \emph {et~al.}(2020)\citenamefont
  {Bonilla}, \citenamefont {D{\textquotesingle}Agostino}, \citenamefont
  {Nunes},\ and\ \citenamefont {de~Araujo}}]{Bonilla_2020}%
  \BibitemOpen
  \bibfield  {author} {\bibinfo {author} {\bibfnamefont {A.}~\bibnamefont
  {Bonilla}}, \bibinfo {author} {\bibfnamefont {R.}~\bibnamefont
  {D{\textquotesingle}Agostino}}, \bibinfo {author} {\bibfnamefont {R.~C.}\
  \bibnamefont {Nunes}}, \ and\ \bibinfo {author} {\bibfnamefont {J.~C.}\
  \bibnamefont {de~Araujo}},\ }\href {\doibase 10.1088/1475-7516/2020/03/015}
  {\bibfield  {journal} {\bibinfo  {journal} {Journal of Cosmology and
  Astroparticle Physics}\ }\textbf {\bibinfo {volume} {2020}},\ \bibinfo
  {pages} {015} (\bibinfo {year} {2020})}\BibitemShut {NoStop}%
\bibitem [{\citenamefont {Deffayet}\ \emph {et~al.}(2010)\citenamefont
  {Deffayet}, \citenamefont {Pujolas}, \citenamefont {Sawicki},\ and\
  \citenamefont {Vikman}}]{Deffayet:2010qz}%
  \BibitemOpen
  \bibfield  {author} {\bibinfo {author} {\bibfnamefont {C.}~\bibnamefont
  {Deffayet}}, \bibinfo {author} {\bibfnamefont {O.}~\bibnamefont {Pujolas}},
  \bibinfo {author} {\bibfnamefont {I.}~\bibnamefont {Sawicki}}, \ and\
  \bibinfo {author} {\bibfnamefont {A.}~\bibnamefont {Vikman}},\ }\href
  {\doibase 10.1088/1475-7516/2010/10/026} {\bibfield  {journal} {\bibinfo
  {journal} {JCAP}\ }\textbf {\bibinfo {volume} {1010}},\ \bibinfo {pages}
  {026} (\bibinfo {year} {2010})},\ \Eprint {http://arxiv.org/abs/1008.0048}
  {arXiv:1008.0048 [hep-th]} \BibitemShut {NoStop}%
\bibitem [{\citenamefont {Hui}\ \emph {et~al.}(2009)\citenamefont {Hui},
  \citenamefont {Nicolis},\ and\ \citenamefont {Stubbs}}]{Hui:2009kc}%
  \BibitemOpen
  \bibfield  {author} {\bibinfo {author} {\bibfnamefont {L.}~\bibnamefont
  {Hui}}, \bibinfo {author} {\bibfnamefont {A.}~\bibnamefont {Nicolis}}, \ and\
  \bibinfo {author} {\bibfnamefont {C.}~\bibnamefont {Stubbs}},\ }\href
  {\doibase 10.1103/PhysRevD.80.104002} {\bibfield  {journal} {\bibinfo
  {journal} {Phys. Rev.}\ }\textbf {\bibinfo {volume} {D80}},\ \bibinfo {pages}
  {104002} (\bibinfo {year} {2009})},\ \Eprint {http://arxiv.org/abs/0905.2966}
  {arXiv:0905.2966 [astro-ph.CO]} \BibitemShut {NoStop}%
\bibitem [{\citenamefont {Hofmann}\ and\ \citenamefont
  {M{\"u}ller}(2018)}]{Hofmann:2018myc}%
  \BibitemOpen
  \bibfield  {author} {\bibinfo {author} {\bibfnamefont {F.}~\bibnamefont
  {Hofmann}}\ and\ \bibinfo {author} {\bibfnamefont {J.}~\bibnamefont
  {M{\"u}ller}},\ }\href {\doibase 10.1088/1361-6382/aa8f7a} {\bibfield
  {journal} {\bibinfo  {journal} {Class. Quant. Grav.}\ }\textbf {\bibinfo
  {volume} {35}},\ \bibinfo {pages} {035015} (\bibinfo {year}
  {2018})}\BibitemShut {NoStop}%
\bibitem [{\citenamefont {{Zhu}}\ \emph {et~al.}(2015)\citenamefont {{Zhu}},
  \citenamefont {{Stairs}}, \citenamefont {{Demorest}}, \citenamefont {{Nice}},
  \citenamefont {{Ellis}}, \citenamefont {{Ransom}}, \citenamefont
  {{Arzoumanian}}, \citenamefont {{Crowter}}, \citenamefont {{Dolch}},
  \citenamefont {{Ferdman}}, \citenamefont {{Fonseca}}, \citenamefont
  {{Gonzalez}}, \citenamefont {{Jones}}, \citenamefont {{Jones}}, \citenamefont
  {{Lam}}, \citenamefont {{Levin}}, \citenamefont {{McLaughlin}}, \citenamefont
  {{Pennucci}}, \citenamefont {{Stovall}},\ and\ \citenamefont
  {{Swiggum}}}]{2015ApJ...809...41Z}%
  \BibitemOpen
  \bibfield  {author} {\bibinfo {author} {\bibfnamefont {W.~W.}\ \bibnamefont
  {{Zhu}}}, \bibinfo {author} {\bibfnamefont {I.~H.}\ \bibnamefont {{Stairs}}},
  \bibinfo {author} {\bibfnamefont {P.~B.}\ \bibnamefont {{Demorest}}},
  \bibinfo {author} {\bibfnamefont {D.~J.}\ \bibnamefont {{Nice}}}, \bibinfo
  {author} {\bibfnamefont {J.~A.}\ \bibnamefont {{Ellis}}}, \bibinfo {author}
  {\bibfnamefont {S.~M.}\ \bibnamefont {{Ransom}}}, \bibinfo {author}
  {\bibfnamefont {Z.}~\bibnamefont {{Arzoumanian}}}, \bibinfo {author}
  {\bibfnamefont {K.}~\bibnamefont {{Crowter}}}, \bibinfo {author}
  {\bibfnamefont {T.}~\bibnamefont {{Dolch}}}, \bibinfo {author} {\bibfnamefont
  {R.~D.}\ \bibnamefont {{Ferdman}}}, \bibinfo {author} {\bibfnamefont
  {E.}~\bibnamefont {{Fonseca}}}, \bibinfo {author} {\bibfnamefont {M.~E.}\
  \bibnamefont {{Gonzalez}}}, \bibinfo {author} {\bibfnamefont
  {G.}~\bibnamefont {{Jones}}}, \bibinfo {author} {\bibfnamefont {M.~L.}\
  \bibnamefont {{Jones}}}, \bibinfo {author} {\bibfnamefont {M.~T.}\
  \bibnamefont {{Lam}}}, \bibinfo {author} {\bibfnamefont {L.}~\bibnamefont
  {{Levin}}}, \bibinfo {author} {\bibfnamefont {M.~A.}\ \bibnamefont
  {{McLaughlin}}}, \bibinfo {author} {\bibfnamefont {T.}~\bibnamefont
  {{Pennucci}}}, \bibinfo {author} {\bibfnamefont {K.}~\bibnamefont
  {{Stovall}}}, \ and\ \bibinfo {author} {\bibfnamefont {J.}~\bibnamefont
  {{Swiggum}}},\ }\href {\doibase 10.1088/0004-637X/809/1/41} {\bibfield
  {journal} {\bibinfo  {journal} {\apj}\ }\textbf {\bibinfo {volume} {809}},\
  \bibinfo {eid} {41} (\bibinfo {year} {2015})},\ \Eprint
  {http://arxiv.org/abs/1504.00662} {arXiv:1504.00662 [astro-ph.SR]}
  \BibitemShut {NoStop}%
\bibitem [{\citenamefont {Bertotti}\ \emph {et~al.}(2003)\citenamefont
  {Bertotti}, \citenamefont {Iess},\ and\ \citenamefont
  {Tortora}}]{Bertotti:2003rm}%
  \BibitemOpen
  \bibfield  {author} {\bibinfo {author} {\bibfnamefont {B.}~\bibnamefont
  {Bertotti}}, \bibinfo {author} {\bibfnamefont {L.}~\bibnamefont {Iess}}, \
  and\ \bibinfo {author} {\bibfnamefont {P.}~\bibnamefont {Tortora}},\ }\href
  {\doibase 10.1038/nature01997} {\bibfield  {journal} {\bibinfo  {journal}
  {Nature}\ }\textbf {\bibinfo {volume} {425}},\ \bibinfo {pages} {374}
  (\bibinfo {year} {2003})}\BibitemShut {NoStop}%
\bibitem [{\citenamefont {Hinterbichler}\ and\ \citenamefont
  {Khoury}(2010)}]{PhysRevLett.104.231301}%
  \BibitemOpen
  \bibfield  {author} {\bibinfo {author} {\bibfnamefont {K.}~\bibnamefont
  {Hinterbichler}}\ and\ \bibinfo {author} {\bibfnamefont {J.}~\bibnamefont
  {Khoury}},\ }\href {\doibase 10.1103/PhysRevLett.104.231301} {\bibfield
  {journal} {\bibinfo  {journal} {Phys. Rev. Lett.}\ }\textbf {\bibinfo
  {volume} {104}},\ \bibinfo {pages} {231301} (\bibinfo {year}
  {2010})}\BibitemShut {NoStop}%
\bibitem [{\citenamefont {Hinterbichler}\ \emph {et~al.}(2011)\citenamefont
  {Hinterbichler}, \citenamefont {Khoury}, \citenamefont {Levy},\ and\
  \citenamefont {Matas}}]{Hinterbichler:2011ca}%
  \BibitemOpen
  \bibfield  {author} {\bibinfo {author} {\bibfnamefont {K.}~\bibnamefont
  {Hinterbichler}}, \bibinfo {author} {\bibfnamefont {J.}~\bibnamefont
  {Khoury}}, \bibinfo {author} {\bibfnamefont {A.}~\bibnamefont {Levy}}, \ and\
  \bibinfo {author} {\bibfnamefont {A.}~\bibnamefont {Matas}},\ }\href
  {\doibase 10.1103/PhysRevD.84.103521} {\bibfield  {journal} {\bibinfo
  {journal} {Phys. Rev.}\ }\textbf {\bibinfo {volume} {D84}},\ \bibinfo {pages}
  {103521} (\bibinfo {year} {2011})},\ \Eprint {http://arxiv.org/abs/1107.2112}
  {arXiv:1107.2112 [astro-ph.CO]} \BibitemShut {NoStop}%
\bibitem [{\citenamefont {de~Rham}\ \emph {et~al.}(2013)\citenamefont
  {de~Rham}, \citenamefont {Tolley},\ and\ \citenamefont
  {Wesley}}]{deRham:2012fw}%
  \BibitemOpen
  \bibfield  {author} {\bibinfo {author} {\bibfnamefont {C.}~\bibnamefont
  {de~Rham}}, \bibinfo {author} {\bibfnamefont {A.~J.}\ \bibnamefont {Tolley}},
  \ and\ \bibinfo {author} {\bibfnamefont {D.~H.}\ \bibnamefont {Wesley}},\
  }\href {\doibase 10.1103/PhysRevD.87.044025} {\bibfield  {journal} {\bibinfo
  {journal} {Phys. Rev.}\ }\textbf {\bibinfo {volume} {D87}},\ \bibinfo {pages}
  {044025} (\bibinfo {year} {2013})},\ \Eprint {http://arxiv.org/abs/1208.0580}
  {arXiv:1208.0580 [gr-qc]} \BibitemShut {NoStop}%
\bibitem [{\citenamefont {Chu}\ and\ \citenamefont
  {Trodden}(2013)}]{Chu:2012kz}%
  \BibitemOpen
  \bibfield  {author} {\bibinfo {author} {\bibfnamefont {Y.-Z.}\ \bibnamefont
  {Chu}}\ and\ \bibinfo {author} {\bibfnamefont {M.}~\bibnamefont {Trodden}},\
  }\href {\doibase 10.1103/PhysRevD.87.024011} {\bibfield  {journal} {\bibinfo
  {journal} {Phys. Rev.}\ }\textbf {\bibinfo {volume} {D87}},\ \bibinfo {pages}
  {024011} (\bibinfo {year} {2013})},\ \Eprint {http://arxiv.org/abs/1210.6651}
  {arXiv:1210.6651 [astro-ph.CO]} \BibitemShut {NoStop}%
\bibitem [{\citenamefont {Dar}\ \emph {et~al.}(2019)\citenamefont {Dar},
  \citenamefont {De~Rham}, \citenamefont {Deskins}, \citenamefont {Giblin},\
  and\ \citenamefont {Tolley}}]{Dar:2018dra}%
  \BibitemOpen
  \bibfield  {author} {\bibinfo {author} {\bibfnamefont {F.}~\bibnamefont
  {Dar}}, \bibinfo {author} {\bibfnamefont {C.}~\bibnamefont {De~Rham}},
  \bibinfo {author} {\bibfnamefont {J.~T.}\ \bibnamefont {Deskins}}, \bibinfo
  {author} {\bibfnamefont {J.~T.}\ \bibnamefont {Giblin}}, \ and\ \bibinfo
  {author} {\bibfnamefont {A.~J.}\ \bibnamefont {Tolley}},\ }\href {\doibase
  10.1088/1361-6382/aaf5e8} {\bibfield  {journal} {\bibinfo  {journal} {Class.
  Quant. Grav.}\ }\textbf {\bibinfo {volume} {36}},\ \bibinfo {pages} {025008}
  (\bibinfo {year} {2019})},\ \Eprint {http://arxiv.org/abs/1808.02165}
  {arXiv:1808.02165 [hep-th]} \BibitemShut {NoStop}%
\bibitem [{\citenamefont {Katsuragawa}\ \emph {et~al.}(2019)\citenamefont
  {Katsuragawa}, \citenamefont {Nakamura}, \citenamefont {Ikeda},\ and\
  \citenamefont {Capozziello}}]{Katsuragawa:2019uto}%
  \BibitemOpen
  \bibfield  {author} {\bibinfo {author} {\bibfnamefont {T.}~\bibnamefont
  {Katsuragawa}}, \bibinfo {author} {\bibfnamefont {T.}~\bibnamefont
  {Nakamura}}, \bibinfo {author} {\bibfnamefont {T.}~\bibnamefont {Ikeda}}, \
  and\ \bibinfo {author} {\bibfnamefont {S.}~\bibnamefont {Capozziello}},\
  }\href {\doibase 10.1103/PhysRevD.99.124050} {\bibfield  {journal} {\bibinfo
  {journal} {Phys. Rev. D}\ }\textbf {\bibinfo {volume} {99}},\ \bibinfo
  {pages} {124050} (\bibinfo {year} {2019})},\ \Eprint
  {http://arxiv.org/abs/1902.02494} {arXiv:1902.02494 [gr-qc]} \BibitemShut
  {NoStop}%
\bibitem [{\citenamefont {Hou}\ \emph {et~al.}(2018)\citenamefont {Hou},
  \citenamefont {Gong},\ and\ \citenamefont {Liu}}]{Hou:2017bqj}%
  \BibitemOpen
  \bibfield  {author} {\bibinfo {author} {\bibfnamefont {S.}~\bibnamefont
  {Hou}}, \bibinfo {author} {\bibfnamefont {Y.}~\bibnamefont {Gong}}, \ and\
  \bibinfo {author} {\bibfnamefont {Y.}~\bibnamefont {Liu}},\ }\href {\doibase
  10.1140/epjc/s10052-018-5869-y} {\bibfield  {journal} {\bibinfo  {journal}
  {Eur. Phys. J.}\ }\textbf {\bibinfo {volume} {C78}},\ \bibinfo {pages} {378}
  (\bibinfo {year} {2018})},\ \Eprint {http://arxiv.org/abs/1704.01899}
  {arXiv:1704.01899 [gr-qc]} \BibitemShut {NoStop}%
\bibitem [{\citenamefont {Garoffolo}\ \emph {et~al.}(2019)\citenamefont
  {Garoffolo}, \citenamefont {Tasinato}, \citenamefont {Carbone}, \citenamefont
  {Bertacca},\ and\ \citenamefont {Matarrese}}]{Garoffolo:2019mna}%
  \BibitemOpen
  \bibfield  {author} {\bibinfo {author} {\bibfnamefont {A.}~\bibnamefont
  {Garoffolo}}, \bibinfo {author} {\bibfnamefont {G.}~\bibnamefont {Tasinato}},
  \bibinfo {author} {\bibfnamefont {C.}~\bibnamefont {Carbone}}, \bibinfo
  {author} {\bibfnamefont {D.}~\bibnamefont {Bertacca}}, \ and\ \bibinfo
  {author} {\bibfnamefont {S.}~\bibnamefont {Matarrese}},\ }\href@noop {} {\
  (\bibinfo {year} {2019})},\ \Eprint {http://arxiv.org/abs/1912.08093}
  {arXiv:1912.08093 [gr-qc]} \BibitemShut {NoStop}%
\bibitem [{\citenamefont {Weinberg}(2004)}]{Weinberg:2003ur}%
  \BibitemOpen
  \bibfield  {author} {\bibinfo {author} {\bibfnamefont {S.}~\bibnamefont
  {Weinberg}},\ }\href {\doibase 10.1103/PhysRevD.69.023503} {\bibfield
  {journal} {\bibinfo  {journal} {Phys. Rev.}\ }\textbf {\bibinfo {volume}
  {D69}},\ \bibinfo {pages} {023503} (\bibinfo {year} {2004})},\ \Eprint
  {http://arxiv.org/abs/astro-ph/0306304} {arXiv:astro-ph/0306304 [astro-ph]}
  \BibitemShut {NoStop}%
\bibitem [{\citenamefont {Scomparin}\ and\ \citenamefont
  {Vazzoler}(2019)}]{Scomparin:2019ziw}%
  \BibitemOpen
  \bibfield  {author} {\bibinfo {author} {\bibfnamefont {M.}~\bibnamefont
  {Scomparin}}\ and\ \bibinfo {author} {\bibfnamefont {S.}~\bibnamefont
  {Vazzoler}},\ }\href@noop {} {\  (\bibinfo {year} {2019})},\ \Eprint
  {http://arxiv.org/abs/1903.01502} {arXiv:1903.01502 [gr-qc]} \BibitemShut
  {NoStop}%
\bibitem [{\citenamefont {Straumann}(2013)}]{Straumann:2013spu}%
  \BibitemOpen
  \bibfield  {author} {\bibinfo {author} {\bibfnamefont {N.}~\bibnamefont
  {Straumann}},\ }\href {\doibase 10.1007/978-94-007-5410-2} {\emph {\bibinfo
  {title} {{General Relativity}}}},\ Graduate Texts in Physics\ (\bibinfo
  {publisher} {Springer},\ \bibinfo {address} {Dordrecht},\ \bibinfo {year}
  {2013})\BibitemShut {NoStop}%
\bibitem [{\citenamefont {Harte}(2019)}]{Harte:2018wni}%
  \BibitemOpen
  \bibfield  {author} {\bibinfo {author} {\bibfnamefont {A.~I.}\ \bibnamefont
  {Harte}},\ }\href {\doibase 10.1007/s10714-018-2494-x} {\bibfield  {journal}
  {\bibinfo  {journal} {Gen. Rel. Grav.}\ }\textbf {\bibinfo {volume} {51}},\
  \bibinfo {pages} {14} (\bibinfo {year} {2019})},\ \Eprint
  {http://arxiv.org/abs/1808.06203} {arXiv:1808.06203 [gr-qc]} \BibitemShut
  {NoStop}%
\bibitem [{\citenamefont {Cusin}\ and\ \citenamefont
  {Lagos}(2019)}]{Cusin:2019rmt}%
  \BibitemOpen
  \bibfield  {author} {\bibinfo {author} {\bibfnamefont {G.}~\bibnamefont
  {Cusin}}\ and\ \bibinfo {author} {\bibfnamefont {M.}~\bibnamefont {Lagos}},\
  }\href@noop {} {\  (\bibinfo {year} {2019})},\ \Eprint
  {http://arxiv.org/abs/1910.13326} {arXiv:1910.13326 [gr-qc]} \BibitemShut
  {NoStop}%
\bibitem [{\citenamefont {Fleury}(2015)}]{Fleury:2015hgz}%
  \BibitemOpen
  \bibfield  {author} {\bibinfo {author} {\bibfnamefont {P.}~\bibnamefont
  {Fleury}},\ }\emph {\bibinfo {title} {{Light propagation in inhomogeneous and
  anisotropic cosmologies}}},\ \href@noop {} {Ph.D. thesis},\ \bibinfo
  {school} {Paris, Inst. Astrophys.} (\bibinfo {year} {2015}),\ \Eprint
  {http://arxiv.org/abs/1511.03702} {arXiv:1511.03702 [gr-qc]} \BibitemShut
  {NoStop}%
\bibitem [{\citenamefont {{Etherington}}(1933)}]{1933PMag...15..761E}%
  \BibitemOpen
  \bibfield  {author} {\bibinfo {author} {\bibfnamefont {I.~M.~H.}\
  \bibnamefont {{Etherington}}},\ }\href@noop {} {\bibfield  {journal}
  {\bibinfo  {journal} {Philosophical Magazine}\ }\textbf {\bibinfo {volume}
  {15}},\ \bibinfo {pages} {761} (\bibinfo {year} {1933})}\BibitemShut
  {NoStop}%
\bibitem [{\citenamefont {{Schneider}}\ \emph {et~al.}(1992)\citenamefont
  {{Schneider}}, \citenamefont {{Ehlers}},\ and\ \citenamefont
  {{Falco}}}]{1992grle.book.....S}%
  \BibitemOpen
  \bibfield  {author} {\bibinfo {author} {\bibfnamefont {P.}~\bibnamefont
  {{Schneider}}}, \bibinfo {author} {\bibfnamefont {J.}~\bibnamefont
  {{Ehlers}}}, \ and\ \bibinfo {author} {\bibfnamefont {E.~E.}\ \bibnamefont
  {{Falco}}},\ }\href {\doibase 10.1007/978-3-662-03758-4} {\emph {\bibinfo
  {title} {{Gravitational Lenses}}}}\ (\bibinfo {year} {1992})\BibitemShut
  {NoStop}%
\bibitem [{\citenamefont {Maggiore}(2007)}]{Maggiore:1900zz}%
  \BibitemOpen
  \bibfield  {author} {\bibinfo {author} {\bibfnamefont {M.}~\bibnamefont
  {Maggiore}},\ }\href {http://www.oup.com/uk/catalogue/?ci=9780198570745}
  {\emph {\bibinfo {title} {{Gravitational Waves. Vol. 1: Theory and
  Experiments}}}},\ Oxford Master Series in Physics\ (\bibinfo  {publisher}
  {Oxford University Press},\ \bibinfo {year} {2007})\BibitemShut {NoStop}%
\bibitem [{\citenamefont {Wolf}\ and\ \citenamefont
  {Lagos}(2019)}]{Wolf:2019hun}%
  \BibitemOpen
  \bibfield  {author} {\bibinfo {author} {\bibfnamefont {W.~J.}\ \bibnamefont
  {Wolf}}\ and\ \bibinfo {author} {\bibfnamefont {M.}~\bibnamefont {Lagos}},\
  }\href@noop {} {\  (\bibinfo {year} {2019})},\ \Eprint
  {http://arxiv.org/abs/1910.10580} {arXiv:1910.10580 [gr-qc]} \BibitemShut
  {NoStop}%
\bibitem [{\citenamefont {Tsujikawa}(2019)}]{Tsujikawa:2019pih}%
  \BibitemOpen
  \bibfield  {author} {\bibinfo {author} {\bibfnamefont {S.}~\bibnamefont
  {Tsujikawa}},\ }\href@noop {} {\  (\bibinfo {year} {2019})},\ \Eprint
  {http://arxiv.org/abs/1903.07092} {arXiv:1903.07092 [gr-qc]} \BibitemShut
  {NoStop}%
\bibitem [{\citenamefont {{Jackson}}(1998)}]{1998clel.book.....J}%
  \BibitemOpen
  \bibfield  {author} {\bibinfo {author} {\bibfnamefont {J.~D.}\ \bibnamefont
  {{Jackson}}},\ }\href@noop {} {\emph {\bibinfo {title} {{Classical
  Electrodynamics, 3rd Edition}}}}\ (\bibinfo {year} {1998})\BibitemShut
  {NoStop}%
\end{thebibliography}%
\end{document}